\documentclass[fleqn,usenatbib]{mnras}

\usepackage{newtxtext,newtxmath}

\usepackage[T1]{fontenc}

\DeclareRobustCommand{\VAN}[3]{#2}
\let\VANthebibliography\thebibliography
\def\thebibliography{\DeclareRobustCommand{\VAN}[3]{##3}\VANthebibliography}

\usepackage{graphicx}	% Including figure files
\usepackage{amsmath}	% Advanced maths commands
\usepackage{amssymb}	% Extra maths symbols

\title[RSFs as BHNS and NSNS merger counterparts]{Resonant Shattering Flares in Black Hole-Neutron Star and Binary Neutron Star Mergers}

\author[D. Neill et al.]{
Duncan Neill$^{1} $\thanks{E-mail: dn431@bath.ac.uk},
David Tsang$^{1}$, Hendrik van Eerten$^{1}$, 
Geoffrey Ryan $^{2}$, and William G. Newton$^{3}$
\\
$^{1}$Department of Physics, University of Bath, Claverton Down, Bath, BA2 7AY, UK\\
$^{2}$Perimeter Institute for Theoretical Physics, 31 Caroline St. N., Waterloo, ON, N2L 2Y5, Canada\\
$^{3}$Department of Physics and Astronomy, Texas A\&M University-Commerce, Commerce, TX, 75429-3011, USA
}

\date{Accepted XXX. Received YYY; in original form ZZZ}

\pubyear{2022}

\begin{document}
\label{firstpage}
\pagerange{\pageref{firstpage}--\pageref{lastpage}}
\maketitle

\begin{abstract}

Resonant Shattering flares (RSFs) are bursts of gamma-rays expected to be triggered by tidal resonance of a neutron star (NS) during binary inspiral. They are strongly dependent on the magnetic field strength at the surface of the NS. By modelling these flares as being the result of multiple colliding relativistic shells launched during the resonance window, we find that the prompt non-thermal gamma-ray emission may have luminosity up to a few $\times10^{48}\text{ erg/s}$, and that a broad-band afterglow could be produced. We compute the expected rates of detectable RSFs using the BPASS population synthesis code, with different assumptions about the evolution of surface magnetic field strengths before merger. We find the rate of detectable RSFs to be $\sim 0.0001-5$ per year for BHNS mergers and $\sim 0.0005-25$ per year for NSNS mergers, with the lower bound corresponding to surface-field decay consistent with magneto-thermal evolution in purely crustal fields, while the upper bounds are for systems which have longer-lived surface magnetic fields supported by flux frozen into the superconducting core. If some of the observed SGRB precursor flares are indeed RSFs, this suggests the presence of a longer-lived surface field for some fraction of the neutron star population, and that we could expect RSFs to be the most common detectable EM counterpart to GW detections of BHNS mergers. The non-detection of a RSF prior to GRB170817A provides an upper bound on the magnetic fields of the progenitor NSs of $B_{\rm surf}\sim 10^{13.5} \text{ G}$.

\end{abstract}

\begin{keywords}
dense matter -- stars: neutron -- black hole - neutron star mergers -- neutron star mergers -- gravitational waves -- transients: gamma-ray bursts
\end{keywords}

%%%%%%%%%%%%%%%%%%%%%%%%%%%%%%%%%%%%%%%%%%%%%%%%%%

%%%%%%%%%%%%%%%%% BODY OF PAPER %%%%%%%%%%%%%%%%%%

\section{Introduction}

Gravitational-wave (GW) astronomy recently reached another major milestone, with the first two detections of likely Black Hole (BH) - Neutron Star (NS) mergers \citep{abbott2021observation} discovered within a week of one another. This achievement adds to the already impressive history of the nascent field, beginning with the first detection of gravitational-waves (from a BHBH merger) in 2015 \citep{abbott2016observation}, as well as the detection of the first NSNS merger, GW170817 \citep{abbott2017gw170817}.

GW170817 itself was a watershed event for multi-messenger astrophysics, as it was the first detection of a GW signal accompanied by detected electromagnetic counterparts, a (weak/off-axis) Short Gamma Ray Burst (SGRB) and a kilonova (KN). The detection of this electromagnetic emission not only allowed localisation of the event within the GW error region, but also provided a wealth of insight into the extreme physics of compact object mergers. The KN confirmed NSNS mergers as a major production site of heavy elements via rapid neutron capture nucleosynthesis (e.g. \citealt{arcavi2017, chornock2017, cowperthwaite2017, drout2017, kasen2017, kasliwal2017, pian2017, tanvir2017, troja2017xray}), while the prompt GRB and afterglow emission were found to be consistent with a regular SGRB seen off-axis (e.g. \citealt{goldstein2017ordinary, margutti2017, troja2017xray}), thus confirming the long-standing hypothesis that NSNS mergers are at least one of the formation channels of SGRBs. The source was eventually solidly confirmed to be a collimated jet observed at an angle (both by direct modeling of the light curve turnover, \citealt{troja2018, alexander2018, troja2019year}, and very long baseline interferometry, \citealt{mooley2018superluminal, ghirlanda2019compact}), meaning the outflow structure along the edges of the jet was in full view, carrying the imprint of the launching and early propagation stages of the collimated plasma flow.

In all, coincident EM-GW detections allow physics of the most extreme matter in the universe to be probed in unprecedented detail. In this paper we will investigate the emission and detectability of an EM counterpart that, if detected, can provide strong constraints on nuclear physics of neutron star matter \citep{neill2021resonant, tsang2012resonant}: Resonant Shattering Flares.

\subsection{Resonant Shattering Flares}

During the in-spiral of BHNS and NSNS binaries, the orbital frequency of the binary increases. As this frequency approaches the natural frequency of one of the modes of oscillation within the NS, resonant excitation of that mode can occur. If a particular mode is sufficiently excited by the oscillating tidal field of the binary orbit, this resonance can increase the amplitude of the mode's oscillations enough that it deforms the NS crust to the point that it fractures. The energy of the mode will then be transferred to seismic waves in the crust. Multiple fractures can result in enough energy being in the crust that the whole crust shatters, resulting in much higher frequency seismic waves. If this occurs in a NS which has a sufficiently strong magnetic field \citep{tsang2013shattering}, these high frequency waves can couple to it to produce Alfv\'en waves in the magnetosphere and subsequently lead to the emission of a pair-photon fireball \citep[see, e.g., the mechanisms detailed in][]{thompson1995soft,beloborodov2021emission}. Multiple shattering events can lead to multiple fireballs, the interactions between which cause non-thermal synchrotron emission. If enough shattering events occur over the duration of the resonant excitation, this may produce an observable flare: a resonant shattering flare (RSF).

The quadrupole crust-core interface mode ($i$-mode) of neutron stars has been identified as a likely candidate for triggering these RSFs \citep{tsang2012resonant}. In a previous work \citep{neill2021resonant}, we investigated how multi-messenger observation of a RSF and the GW signal from a binary merger could be used to constrain the properties of the NS crust, particularly near the crust-core boundary. These works did not explore the emission mechanisms and detectability of RSFs in detail, and just focused on what could be learned if any were detected.

A few percent of SGRBs possess precursor flares, some or all of which may be RSFs, as their timings and fluxes relative to their main flares are similar to what simple RSF predictions would suggest. Additionally, as RSFs are thought to be relatively isotropic, we would expect detectable RSFs to be produced by some mergers where the SGRB jet is not directed towards the observer. These ``orphan'' RSFs would then appear as extremely short and weak SGRBs \citep{mandhai2018rate}.

Not all binary mergers produce SGRBs: BHNS mergers can only do so if the NS is tidally disrupted during the in-spiral \citep{Foucart2020}. 
As RSFs do not require this disruption, we expect that they could occur during most BHNS mergers for which the NS has a strong surface magnetic field. Therefore (if NS surface magnetic fields are long-lived) orphan RSFs could be important counterparts to GW observations of BHNS mergers. The first two GW signals with properties consistent with those expected for BHNS mergers were recently detected by the LIGO and VIRGO GW detectors \citep{abbott2021observation}. Neither of these events had coincident EM detections, which matches expectations that tidal disruption is uncommon due to its particular dependence on the properties of the BH and NS \citep[see, e.g.,][]{foucart2012black,fragione2021black}, and we will show that both of these mergers were likely too distant for RSFs to be detectable.

The low rate of BHNS mergers, coupled with the rarity of cases in which tidal disruption occurs, makes multi-messenger observation of these events (and thus more accurate parameter estimation) difficult. In this paper, we will investigate RSFs -- which do not require tidal disruption -- as potential EM counterparts to BHNS (and NSNS) mergers, considering (under various assumptions) the fraction of mergers which do not tidally disrupt, but reach the resonance condition for RSFs  (Section~\ref{sec:td+tr}), RSF luminosities (Section~\ref{sec:RSF_emission}), and their rate in the local universe (Section~\ref{sec:detect}), and compare these with current and future observational prospects.

\section{Tidal Resonance vs Tidal Disruption during BHNS Mergers}\label{sec:td+tr}
\begin{figure*}
\centering
\includegraphics[width=1.0\textwidth,angle=0]{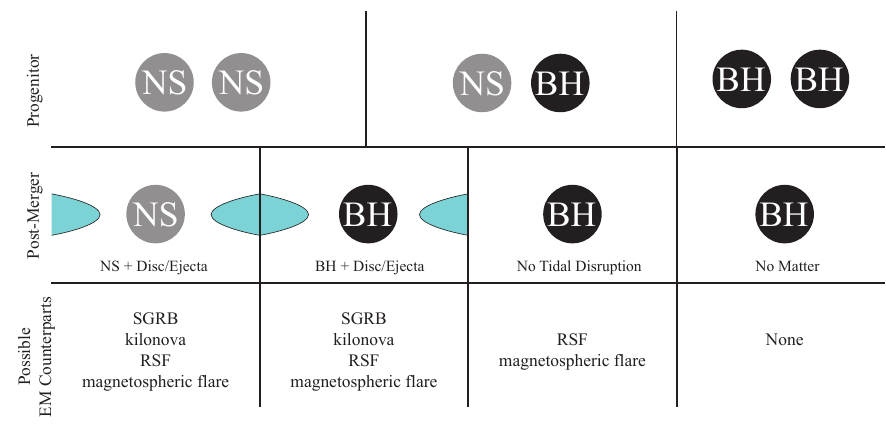}
\caption{Summary table of progenitors, post-merger products, and possible EM counterparts. NSNS and BHNS Mergers can result in SGRBs, KNe, RSFs, and magnetospheric flares \citep[e.g.][]{hansen2001radio,mingarelli2015fast}. However, if no tidal disruption occurs then only RSFs and much weaker flares from magnetospheric interaction may be observed. (Note that here we have ignored EM counterparts from BHBH mergers that have somehow maintained remnant discs, or the AGN disc merger channel.)}
\label{fig:counterparts}
\end{figure*}

In Figure~\ref{fig:counterparts} we show a schematic of the possible EM counterparts for different types of compact object mergers. In this section we shall discuss the requirements for these EM counterparts to be produced (tidal disruption for SGRBs and KNe and tidal resonance for RSFs), and use the results of population synthesis to estimate the fractions of BHNS mergers which satisfy these requirements. 

\subsection{EM counterparts that require BHNS tidal disruption}

The multi-messenger observation of GW170817 and GRB170817A has provided strong evidence that some or all SGRBs originate from NSNS and BHNS mergers. However, the mechanisms behind the emission of a SGRB are complex, with many properties of these bursts being areas of active research (see e.g. \citealt{kumarzhang2015} for a review). It is expected that SGRBs require an accretion disc to be formed following contact between two NSs or tidal disruption of a NS by a BH. This disc can then produce an ultra-relativistic wind via neutrino emission~\citep{mochkovitch1993gamma}, interact with the merger remnant's magnetic field~\citep{blandford1977black}, or in some other way lead to the emission of a SGRB. SGRBs are highly collimated, with the vast majority of the prompt emission occurring within a small angle of the centre of the jet \citep{berger2014, fong2015}. This means that SGRB prompt emission can only be viewed if observed relatively on-axis, and so the SGRBs that are detected may represent only a small fraction of BHNS and NSNS mergers in which tidal disruption occurs. At late times, the jets of SGRBs may spread out as they collide with the ISM, meaning that their afterglows are visible from wider angles (e.g. \citealt{vaneerten2011}). However, without seeing the prompt emission of a SGRB, it would be difficult to know where to look for any orphan afterglows without some other shorter-timescale counterpart, such as a KN or GW localisation.

Kilonovae also require tidal disruption (or contact between NSs for binary NS mergers) in order to eject the required mass from the system \citep{Fernandez2016}. This can be directly ejected by the merger process, or ejected by the remnant disc wind. In either case, a KN will not occur if the NS is swallowed whole by a BH. Unbound neutron-rich ejecta is able to undergo r-process nucleosynthesis, and the radioactive decay of the produced nuclei powers these events~\citep{metzger2010electromagnetic}. However, the KN associated with GW170817 would require $\sim10^{-2} \text{ M}_{\odot}$ of ejecta to match observations \citep[][]{ascenzi2019luminosity, Radice2018Ejection}, which is somewhat higher than predicted by simulations \citep[see e.g. ][]{Hotokezaka2013}, leading some to suggest alternate mechanisms for powering the KN observed after GW170817 \citep[e.g.][]{Metzger2018Magnetar}.

For our purposes, it suffices to note that in order for KNe to be possible for BHNS systems, tidal disruption must first occur. KNe are expected to be more isotropic than SGRBs~\citep{metzger2019kilonovae}, and therefore potentially are detectable as counterparts to some mergers for which we observe no SGRB. However, being significantly dimmer than SGRBs they are not as easily visible from further away, particularly with poor GW localisation \citep{Kasliwal2014}.

\subsection{The properties of merging BHNS systems}

\subsubsection{Population Synthesis}\label{sec:BPASS}
In order to calculate the fraction of BHNS mergers that can produce RSFs, we take the end products of stellar evolution calculated by version 2.2 of the BPASS binary population synthesis code~\citep{eldridge2017binary,stanway2018reevaluating}. BPASS is a population synthesis code designed for the study of the evolution of binary stars. From the results of this population synthesis we take: the number of each of BPASS's BHNS or NSNS systems that are produced per unit mass of star formation at a given metallicity, the masses of the BHs and NSs in these systems, the mass accreted onto the first compact object to form during the evolution of the secondary star, and the orbital separation of the stars at the end of stellar evolution.

As we are interested in the current BHNS merger rate, we only consider mergers that occur within the Hubble time. The time for binary stars to merge when orbital energy is only removed from the system as gravitational waves is given by~\citep{shapiro1983black}
\begin{align}
\tau_0=\frac{5}{256}\frac{c^5}{G^3M_1M_2(M_1+M_2)}r^4,
\label{eq:merge_time}
\end{align}
\noindent where $M_1$ and $M_2$ are the masses of the stars, and $r$ is their separation. This is only an approximation, as it does not consider the impact of eccentricity on the merge times. For a given system, if $\tau_0$ is greater than the Hubble time when both stars have finished their evolution and collapsed to become compact objects, we ignore that system. We do not include the time between star formation and collapse, as it is usually small when compared to $\tau_0$.

BPASS calculates the results of stellar evolution at several different metallicity values. In order to find the weightings given to systems formed at different metallicities in the population of current mergers, we must calculate the amount of star formation that occurred in metallicity bins around each of BPASS's values
at the time each system would need to have formed in order for their mergers to be observed at present. At the time at which a system that is observed to merge at redshift $z_{\rm obs}$ would have formed, the cosmological conditions matched those currently observed at redshift 
\begin{align}
z \approx \left(\left(\frac{\Omega_m}{\Omega_{\Lambda}}\right)^{\frac{1}{3}}\sinh^\frac{2}{3}{\left(\frac{3}{2}H_0\sqrt{\Omega_{\Lambda}}\left(t(z_{\rm obs})-\tau_0\right)\right)}\right)^{-1}-1,
\label{eq:redshift}
\end{align}
\noindent where we have assumed a simplified $\Lambda$CDM cosmology with $H_0\approx 70 \text{ (Km/s)/Mpc}$, $\Omega_m=0.3$, and $\Omega_{\Lambda}=0.7$. $t(z_{\rm obs})$ is the age of the universe when the merger occurred, which is calculated using the inverse of this equation (without $\tau_0$) for $z=z_{\rm obs}$, such that $t(z_{\rm obs})-\tau_0$ is the age of the universe at the time of the second collapse. Again, we ignore the time between the initial star formation and the second collapse, as it is likely small compared to the merge time. The redshift calculated with equation~\eqref{eq:redshift} can be used in the star formation rate history of \citet{madau2014cosmic} 
\begin{align}
\psi(z)=0.015\frac{(1+z)^{2.7}}{1+\left(\frac{(1+z)}{2.9}\right)^{5.6}}\hspace{5pt}\text{ M}_{\odot}\text{ yr}^{-1}\text{ Mpc}^{-3}
\label{eq:SFR}
\end{align}
and the metallicity evolution of \citet{langer2006collapsar} 
\begin{align}
\Psi\left(\frac{Z}{Z_{\odot}},z\right)=\frac{\hat{\Gamma}\left(0.84,\left(\frac{Z}{Z_{\odot}}\right)^210^{0.3z}\right)}{\Gamma(0.84)}
\label{eq:met_frac}
\end{align}
\noindent (where $\Gamma$ and $\hat{\Gamma}$ are the complete and incomplete gamma function) to obtain the amount of star formation occurring at the correct metallicity at the time at which each of BPASS's systems would had to have formed in order to be seen merging at a particular redshift. This, combined with the number of each BPASS system produced per unit mass of star formation, gives us the weighting of each BPASS system in the population of current BHNS mergers.

The set of BPASS results that we have used defines BHs as compact remnants with mass $>3.0 \text{ M}_{\odot}$, and NSs as ones with mass $1.38$ to $3.0 \text{ M}_{\odot}$. Note that objects below $1.38 \text{ M}_{\odot}$ are considered by BPASS to be WDs, and while NSs can have masses below this value, extending the population synthesis to consider this is beyond the scope of this work. BPASS allows super-Eddington accretion to occur onto BHs but forbids it for NSs, increasing the accretion rate of remnants defined as BHs in close binaries (where the Eddington limit is reached). However, BPASS does not produce a mass gap between BHs and NSs, and so there is no obvious distinction between them at $3.0 \text{ M}_{\odot}$, or at any other mass. The choice of $3.0 \text{ M}_{\odot}$ conflicts with current predictions of the NS maximum mass, $M_{\rm max, TOV}\approx 2.0-2.3 \text{ M}_{\odot}$ \citep[see, e.g.][]{margalit2017constraining,rezzolla2018using}, and therefore a rather extreme NS EOS is required to reach $M_{\rm max, TOV}=3.0 \text{ M}_{\odot}$. The NS EOS determines the maximum NS mass and the NS mass-radius relationship (or equivalently, the mass-compactness relationship), and thus the tidal disruption and tidal resonance conditions. We choose to use one NS EOS which results in $M_{\rm max, TOV}= 2.2 \text{ M}_{\odot}$, and another where $M_{\rm max, TOV}= 3.0 \text{ M}_{\odot}$. The first of these EOSs provides a more reasonable NS mass-radius relationship, but is unusable for systems calculated by BPASS where the NS has mass $>2.2 \text{ M}_{\odot}$. Due to the possible differences in accretion, we can not just re-define these NSs as BHs when using this EOS. However, these systems make up only $0.2$\% of all BHNS systems calculated by BPASS. As other uncertainties have a much greater impact on the BHNS population (such as the initial mass function of stars or the dynamics of common envelope phases), we choose to simply ignore these systems. 
BHNS systems in which the BH was born as a NS and underwent accretion induced collapse (AIC) may also have slightly lower mass BHs than if BPASS used $M_{\rm max, TOV}=2.2 \text{ M}_{\odot}$, as they would have been able to undergo super-Eddington accretion when they were between $2.2$ and $3.0 \text{ M}_{\odot}$. However, this is likely to only have a small impact on a small fraction of AIC BHs, and therefore the overall BH mass distribution will not be significantly affected. The second EOS we use produces an unrealistically stiff mass-radius relationship in order to reach $3.0 \text{ M}_{\odot}$. We only consider it to maintain consistency with BPASS, and to show the impact of the stiffness of the NS EOS on our results.

For a description of the model used for the NS EOSs, see \citet{neill2021resonant,newton2020nuclear} and references therein. Note that while in the former the EOS model is described in the context of fixing the maximum mass to $2.2 \text{ M}_{\odot}$, the same model can be used to target $3.0 \text{ M}_{\odot}$ instead. The values of the EOS parameters used for our two EOSs are slightly different, with the $M_{\rm max, TOV}=3.0 \text{ M}_{\odot}$ EOS having lower $L$ and higher $K_{\rm sym}$ parameter values than the $2.2 \text{ M}_{\odot}$ EOS. This is to reduce the stiffness of the $3.0 \text{ M}_{\odot}$ EOS in order to satisfy the tidal deformability bound $\Lambda_{1.4}<800$ calculated from GW170817 \citep{abbott2017gw170817} while having such a high maximum mass. These values conflict with the PREX-II result, which predicts high $L$ \citep{reed2021nuclear, PREXII2021}, and therefore we choose to not use the same low $L$ value for the more reasonable $2.2 \text{ M}_{\odot}$ EOS. However, the large change in $M_{\rm max, TOV}$ from $2.2$ to $3.0 \text{ M}_{\odot}$ has a far more significant impact on EOS stiffness than changes to the other EOS parameters, so our $3.0 \text{ M}_{\odot}$ EOS is much stiffer than our $2.2 \text{ M}_{\odot}$ EOS. We can therefore compare the results for our two EOSs to see the impact of EOS stiffness on tidal disruption and resonance.

\subsubsection{Upper and Lower bounds for Black Hole Spins}
A massive star will have its angular momentum transported to its outer layers if magnetic torques play a significant role in its evolution \citep{Perna2014}. Therefore, when such a star undergoes a supernova, the majority of its angular momentum may be removed as the outer layers are ejected. This can leave natal BHs with very little angular momentum, unless a significant amount is retained by the fallback material. However, a BH with a binary companion that is still burning can be spun up by accretion. The upper limit on the angular momentum transferred to the BH per unit mass accreted corresponds to the situation where all accretion occurs through a thin disc in the BH equatorial plane, with the disc extending to the innermost stable circular orbit (ISCO) and rotating in the same direction as the BH. In this case, the dimensionless spin of the BH $\left(a=\frac{Jc}{GM^2}<1\right)$ can be calculated as~\citep{bardeen1970kerr,thorne1974disk} 
\begin{align}
a=\sqrt{\frac{2}{3}}\frac{M_i}{M}\left(4-\sqrt{18\frac{M_i^2}{M^2}-2}\right),
\label{eq:ang_mom_ulimit}
\end{align}
\noindent where $M_i$ is the initial BH mass (before accretion) and $M$ is its mass after all accretion has occurred. We do not consider natal BH spins from asymmetric supernovae and fallback matter, since these natal spin contributions are small compared to the accretion contribution in this extremely optimistic scenario.

In reality, most accretion occurs during a common envelope phase, where the BH's binary partner expands so that its outer layers encompass the BH \citep{oshaughnessy2005bounds}. Accretion in the common envelope phase is complex, but is certainly much less efficient for angular momentum transfer than a thin disc. Due to the complexity of this system and its inefficiency for angular momentum transfer, we simply use zero as the lower bound on BH angular momentum. We expect that the final spin of the BH after a common envelope phase will be closer to this lower bound than to the extremely optimistic upper bound.

In a small number of the BHNS systems predicted by BPASS, mass transfer between the stars before collapse leads to the less massive NS forming before the BH. In these systems, no accretion onto the BH occurs (except perhaps fallback after its supernova). However, tidal interactions can cause the BH progenitor to spin-up, and if the outer layers of the star have already been lost during accretion, this angular momentum might not be lost when the star collapses. As an upper bound on the BH spin in these systems, we use the formula of \citet{belczynski2020evolutionary}:
\begin{align}
a=e^{-0.1\left(\frac{P_{\rm orb}}{P_0}-1\right)^{1.1}}+0.125,
\label{eq:spin_tidal}
\end{align}
\noindent where $P_0=4000 \text{ s}$ and $0.1\leq P_{\rm orb}\leq 1.3 \text{ days}$ is the orbital period. For $P_{\rm orb}<0.1 \text{ days}$, $a$ is set to $1$, and for $P_{\rm orb}>1.3 \text{ days}$, $a=0$.

\subsection{Tidal Resonance vs Tidal Disruption}\label{sec:td_tr}
Now that we have the masses of the BHs and NSs, the spins of the BHs, and the weightings of the BPASS systems in the population of BHNS mergers, we can calculate the fractions of mergers in which tidal disruption and tidal resonance occur. We calculate the requirement for tidal disruption to occur by following \citet{foucart2012black} and using the fit
\begin{align}
M_{\rm BH}\lesssim M_{\rm NS}\left(R^*_{\rm ISCO}\frac{2.14}{6(C_{\rm NS}^{-1}-2)}\right)^{-1.5},
\label{eq:tid_dis_req}
\end{align}
\noindent where $C_{\rm NS}=\frac{GM_{\rm NS}}{R_{\rm NS}c^2}$ is the compactness of the NS, and $M_{\rm BH}$ and $M_{\rm NS}$ are the masses of the BH and NS. $R^*_{\rm ISCO}$ is the radius of the ISCO in units of the BH mass \citep{bardeen1972rotating}, which introduces a dependence on the BH spin. We assume that the binary orbit is prograde and aligned with respect to the BH spin.

Given a particular spin, we can also estimate an upper bound of $M_{BH}$ for which $i$-mode tidal resonance can occur. This is done by requiring that the ISCO orbital frequency \citep[approximated by, e.g., equation (2.16) of][]{bardeen1972rotating} be greater than half the quadrupole $i$-mode frequency ($f_{2i}$). Rearranging for BH mass, this approximate boundary is given by
\begin{align}
M_{\rm BH} \lesssim \frac{c^3}{G\pi f_{2i}}\left(R_{\rm ISCO}^{* \;\;\;\;\;\; 1.5}+a\right)^{-1},
\label{eq:RSF_req}
\end{align}
\noindent where we use the relativistic NS mode equations of \citet{yoshida2002nonradial} to calculate $f_{2i}\approx 130 \text{ to } 100\text{ Hz}$ for our $M_{\rm max}=2.2\text{ M}_{\odot}$ EOS and $f_{2i}\approx 270 \text{ to } 180\text{ Hz}$ for our $M_{\rm max}=3.0\text{ M}_{\odot}$ EOS, with the ranges being for NS mass varying from $1.4 \text{ M}_{\odot} \text{ to } M_{\rm max}$ \citep[for further details of the $i$-mode, see][]{tsang2012resonant,neill2021resonant}. While this approximation of the orbital frequency at plunge is only valid for extreme mass ratio systems, it suffices to provide a rough estimate for the limit of tidal resonance in the parameter space. As expected, the resonance condition is satisfied for a wider range of BH masses and spins than the tidal disruption condition.

\begin{figure}
\centering
\includegraphics[width=0.49\textwidth,angle=0]{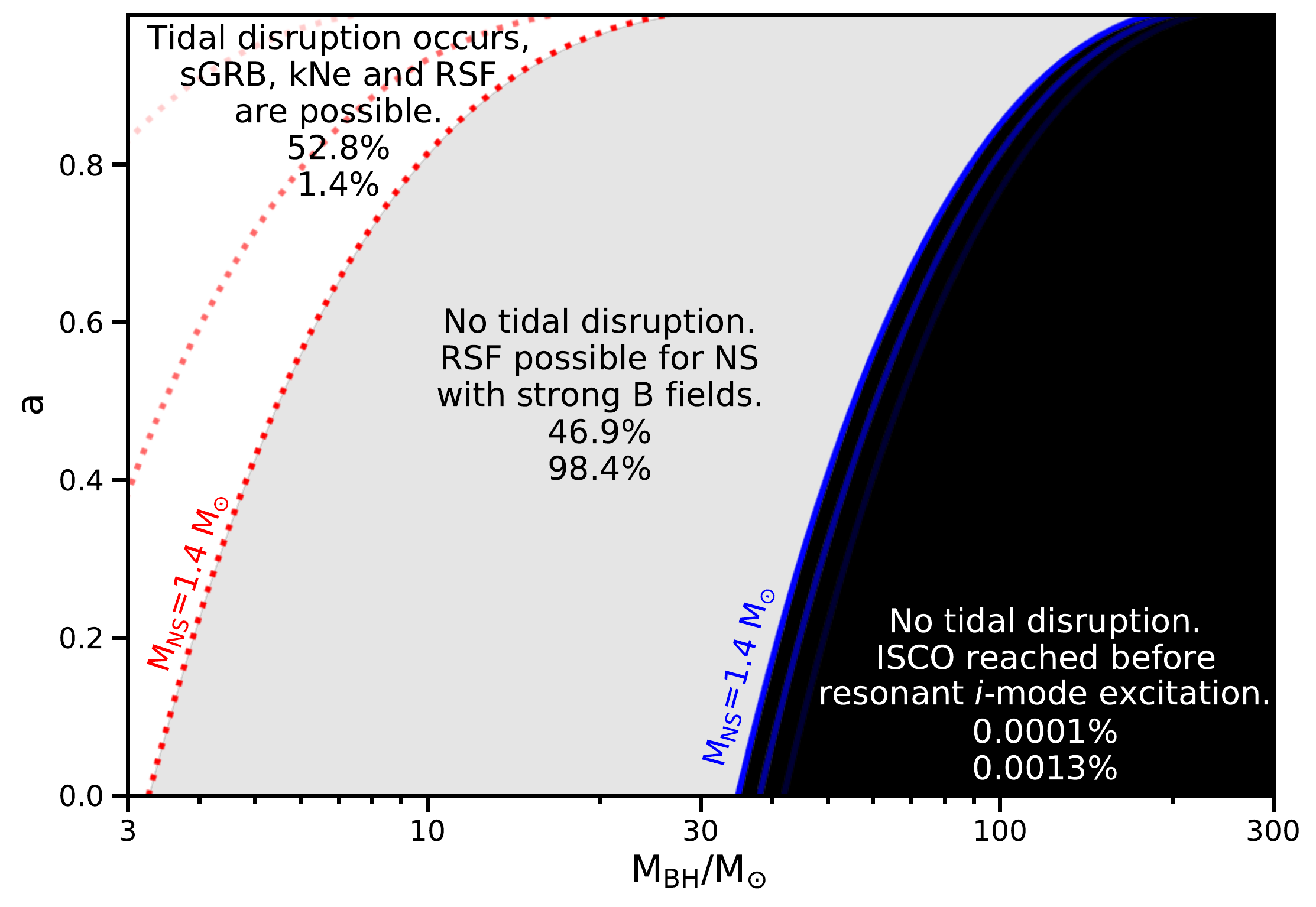}
\caption{The mass-spin plot for BHs in BHNS binaries, showing the conditions for tidal disruption (red dotted line) and tidal resonance (blue line). The lines fade out for higher NS masses, with the first lines being for $1.4 \text{ M}_{\odot}$ and the others for $1.8$ and $2.2 \text{ M}_{\odot}$. These conditions split the plot into three regions: one where SGRBs, KNe, and RSFs can all occur, one where RSFs can occur but SGRBs and KNe can't, and one where no EM counterparts are produced. We show the percentages of BHNS mergers that are in each region, with the upper and lower values being for the upper and lower bounds on BH spin-up due to accretion, respectively. The NS EOS used here (described in \citet{neill2021resonant}) is parameterised by $J=30 \text{ MeV}$, $L=50 \text{ MeV}$, $K_{\rm sym}=-80 \text{ MeV}$, $I_{1.4}=I_{\rm mid}$, and $M_{\rm NS,max}=2.2 \text{ M}_{\odot}$.}
\label{fig:tid_dis_2.2}
\end{figure}

In Figure \ref{fig:tid_dis_2.2} we plot the tidal disruption and tidal resonance conditions for BHNS binaries. These conditions, which depend on BH mass, BH spin, NS mass and NS EOS, separate the plot into three regions. The lines dividing the shaded regions on this plot are for $1.4 \text{ M}_{\odot}$ NSs, while the faded lines are for higher masses, showing that tidal disruption is more difficult and that the in-spiral extends to higher frequencies for more massive NSs. The percentages of BPASS's BHNS population that fall into each of the three regions are shown on the plot, with the upper numbers being for the upper bound on BH spin calculated with equations~\eqref{eq:ang_mom_ulimit} and~\eqref{eq:spin_tidal}, and the lower numbers for the zero-spin lower bound. The $0.2$\% of systems in BPASS's BHNS population that have $M_{\rm NS}>2.2 \text{ M}_{\odot}$ are not included here as their masses can not be produced with this EOS. However, we expect that high-mass NSs such as these would be very compact, and thus unlikely to tidally disrupt. The percentage of systems that tidally disrupt is only high if we are optimistic with the BH spin, and the more realistic lower limit suggests that tidal disruption may be far less common than tidal resonance. Therefore, we find that tidal resonance is likely to be significantly more common than tidal disruption in BHNS mergers, presenting more opportunities for RSFs to be produced than SGRBs or KNe.

\begin{figure}
\centering
\includegraphics[width=0.49\textwidth,angle=0]{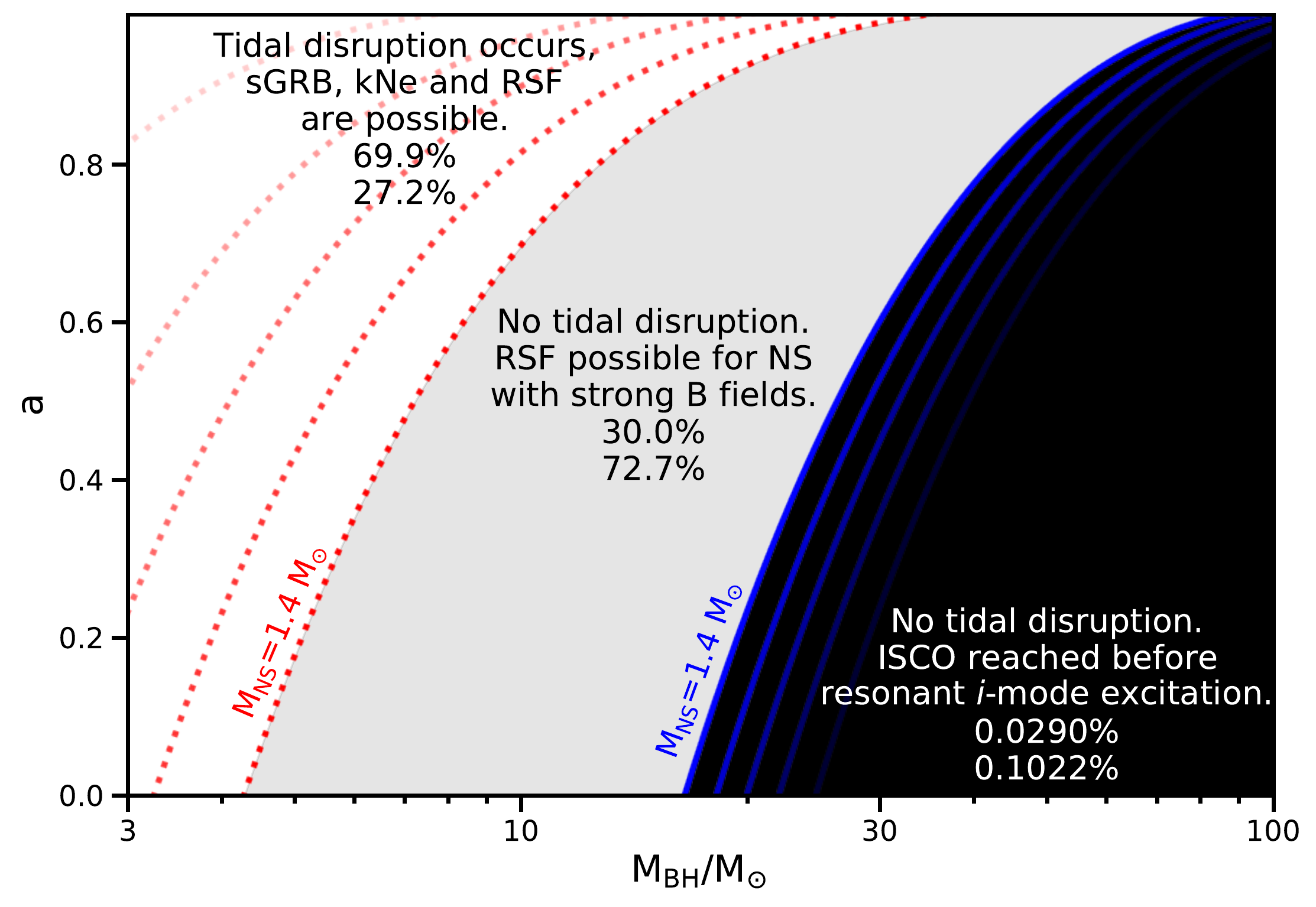}
\caption{Similar to Figure \ref{fig:tid_dis_2.2}, but for an EOS parameterised by $J=31 \text{ MeV}$, $L=39 \text{ MeV}$, $K_{\rm sym}=120 \text{ MeV}$, $I_{1.4}=I_{\rm mid}$, and $M_{\rm NS,max}=3.0 \text{ M}_{\odot}$. This stiffer EOS allows us to include BPASS BHNS systems with $1.4<M_{\rm NS}<3.0 \text{ M}_{\odot}$, and the tidal disruption and resonance conditions are shown for masses in this range (in steps of $0.4 \text{ M}_{\rm NS}$, fading for higher masses).}
\label{fig:tid_dis_3.0}
\end{figure}

For consistency with BPASS's distinction between NSs and BHs, we also calculate tidal disruption and resonance for an EOS with $M_{\rm max,TOV}=3.0 \text{ M}_{\odot}$, shown in Figure \ref{fig:tid_dis_3.0}. While this EOS may be unrealistic it is useful to see that, even for a very stiff EOS (which favours tidal disruption), the low-spin limit still suggests that tidal resonance may be considerably more common than tidal disruption ($99.9$\% vs $27.2$\%).

\section{Emission Mechanisms of RSFs}\label{sec:RSF_emission}
Shattering of the NS crust via resonant excitation of the crust-core interface mode will lead to high-frequency seismic waves in the crust, which subsequently drives Alfv\'en waves in the star's magnetic field. At some distance from the surface of the NS - where the magnetic field is of similar strength to the Alv\'en wave perturbations - interactions between Alfv\'en waves, magnetic reconnection, and other processes are capable of extracting energy from the magnetic field \citep[see, e.g.,][]{thompson1995soft,beloborodov2021emission}. If the extraction of energy is rapid, the area around the emission site will be optically-thick to pair production, resulting in the launch of a pair-photon fireball shell. Multiple fireball shells may be produced within the resonance window.

While for pure dipole fields we may expect Alfv\'en waves to become nonlinear, and thus fireballs to be launched from, far above the NS crust \citep{beloborodov2021emission}, realistic NS magnetic fields likely possess higher-order multipoles \citep{gourgouliatos2016magnetic,braithwaite2008non} which dominate the local magnetic energy density near the NS surface. 
These higher-order multipoles drop off with radius much faster than simple dipole fields, allowing fireballs to be emitted relatively close to the NS crust. 
Since we expect this emission to be dependent on the more complicated multi-pole structure of the magnetic field, it is likely to be relatively isotropic (certainly when compared to SGRBs, which are tightly jetted). We therefore assume that the fireball shells are approximately spherical, and that they are launched from just above the NS crust, resulting in RSFs being isotropic.

In this section we shall explore emission mechanisms by which energy within fireball shells may be released as RSFs.
We focus on the non-thermal prompt emission produced by internal shocks and the afterglow produced by external shocks against the ISM. 
RSF fireball shells will also produce thermal emission, which can be approximated as black-body emission at the temperature at which they are launched \citep{goodman1986gamma}. However, we will show that, with sufficient mass loading, the majority of the energy within RSF shells will be kinetic before they become optically-thin, and so within this model we can assume that thermal emission is negligible.

\subsection{Non-thermal prompt emission from colliding shells emitted during the resonance window}\label{sec:prompt_shell}

Collisions between multiple fireball shells launched during resonance create internal shocks within the shells, which leads to non-thermal synchrotron emission. We follow the method of \citet{kobayashi1997shells} to calculate the luminosity of this prompt emission. This requires us to obtain the initial structure of the shell system, with the important properties being: the number of shells, the distances between shells, the shell thicknesses, the energy of the shells, and the mass loading of the shells.

The energy required to shatter the NS crust is approximated as \citep{tsang2012resonant}
\begin{align}
E_{\rm elastic}\sim2\times10^{46} \text{erg}.
\label{eq:shell_e}
\end{align}
\noindent This is the energy that is eventually emitted as a fireball shell, and so we assume that each shell has energy $E_i\approx E_{\rm elastic}$ (where $i$ indicates the $i$-th shell to be emitted). 
The mass loading of the shells is highly uncertain. We assume that the shells are emitted from close to the NS crust, such that they strip off a fraction of the NS's fluid surface ocean (which has mass $\sim10^{23} \text{ g}$) as they are emitted, and that the ocean is partially replenished between shells. In practice, this means that we assume that the first shell has mass $m_0=10^{23} \text{ g}$, and that all subsequent shells have mass in the range $m_i=10^{21-23} \text{ g}$, with the value for each shell being randomly chosen with flat probability in log-space. The shell energies and masses can be used to calculate their Lorentz factors and velocities as
\begin{align}
\gamma_i=\frac{E_i}{m_ic^2}+1,
\label{eq:shell_lor}
\end{align}
\begin{align}
v_i=\sqrt{c^2-\frac{c^2}{\gamma^2}},
\label{eq:shell_vel}
\end{align}
\noindent with the chosen masses resulting in Lorentz factors ranging from $\sim10^2$ to $\sim10^4$. It is important to note that these masses (and therefore Lorentz factors) have been chosen somewhat arbitrarily, as we have no information on the dynamics involved in the emission of RSFs. However, the shocks that form between shells and the efficiency at which they convert kinetic energy to photons are dependent on the ratios of the shells' Lorentz factors, not their magnitudes. 
Thus, the results of the luminosity calculation are relatively insensitive to the details of the shell masses, so long as there is significant variation between shells, with $\gamma_i \gg 1$. We shall discuss the impact of different shell mass distributions further in Section~\ref{sec:shell_choices}.

There are three notable timescales involved in setting up the system of shells: the resonance timescale $t_{\rm res}$ (the time over which resonant mode excitation and thus shell emission occurs), the breaking timescale $t_{\rm break}$ (the time taken for resonant $i$-mode excitation to shatter the crust), and the emission timescale $t_{\rm emit}$ (the time taken for seismic waves in the crust to be converted into a fireball shell). We estimate the shell thicknesses as $dr_i=v_it_{\rm emit}$, and the number of shells as 
\begin{align}
N=\frac{t_{\rm res}+t_{\rm emit}}{t_{\rm break}+t_{\rm emit}},
\label{eq:num_shells}
\end{align}
\noindent where we have assumed that the shattered crust must heal by fully emitting a shell before resonant excitation can continue. The $t_{\rm emit}$ in the numerator allows the final shell to be emitted outside of the resonance window so long as the crust shattered within it. Finally, the distance between the rear of a shell and the front of next one at the time it begins to be emitted is $\Delta r_{i,i+1}=v_it_{\rm break}$.

These three timescales have significant impacts on the results. We approximate the resonance and breaking timescales as $t_{\rm res}\sim0.1 \text{ s}$ and $t_{\rm break}\sim0.001 \text{ s}$ \citep{tsang2012resonant}. The emission timescale is determined as the time taken for the NS's magnetic field to extract the seismic energy from the crust, $t_{\rm emit}\approx\frac{E_{\rm elastic}}{L_{\rm max}}$. From \citet{tsang2013shattering} we take the rate at which energy is extracted,
\begin{align}
L_{\rm max}\sim10^{47}\text{erg/s}\hspace{5pt}\left(\frac{v}{c}\right)\left(\frac{R_{\rm NS}}{10 \text{Km}}\right)^2\left(\frac{B_{\rm surf}}{10^{13}\text{G}}\right)^2,
\label{eq:emission_rate}
\end{align}
\noindent where $B_{\rm surf}$ is the NS's total surface magnetic field strength (which is the most important parameter that controls the average luminosity of a RSF), and we take the maximum velocity of the perturbation of the field line $v\sim c$.

Now that we have the initial properties of the shells, we follow the method of \citet{kobayashi1997shells} to calculate the luminosity of the emission from the collisions between them. Starting from the time at which the first shell is emitted, 
we determine which shells collide next, and then advance the time of the system to when that collision occurs by increasing the radius of each shell by the distance it travels in that time. 
Assuming that collisions are inelastic, the kinetic energy dissipated in collisions (i.e., the energy of the flare) and the Lorentz factors of the resulting merged shells are calculated by conservation of momentum. 
The profile of the observed burst luminosity is dependent on the thickness of the shells (which determines the time the shocks take to cross them) and the radius at which the collision occurs (which determines angular spread of the components of the shells that contribute to the luminosity along a particular sight line). Once the luminosity of a collision has been calculated, we find the pair of shells which collide next and repeat the process. This continues until no more collisions occur, i.e. when the shells are ordered so that velocity always increases with radius.

\begin{figure}
\centering
\includegraphics[width=0.49\textwidth,angle=0]{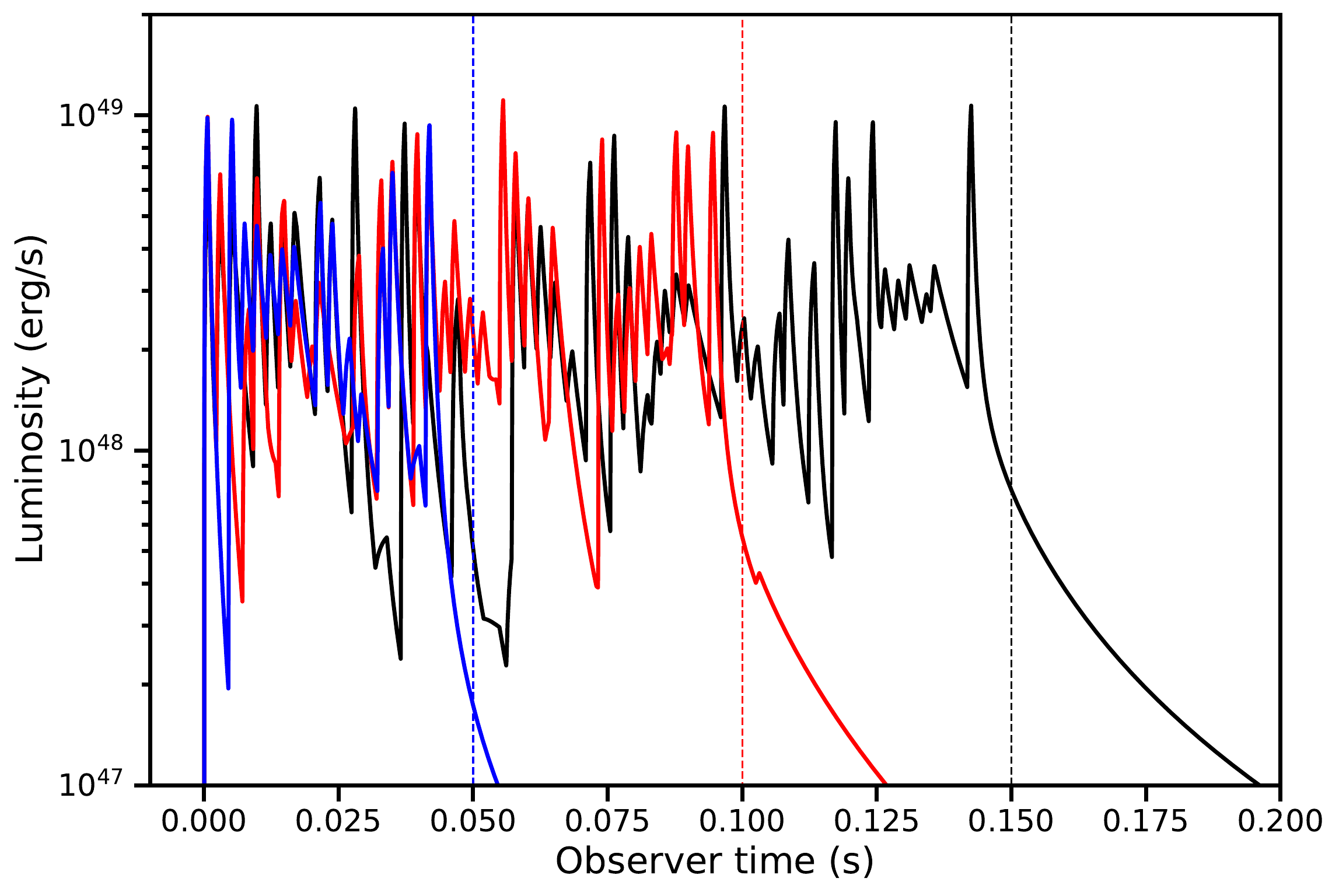}
\caption{The luminosity of the RSF produced by NSs with different resonance timescales. The blue, red and black lines are for $t_{\rm res}=0.05,0.10,0.15 \text{ s}$ (as indicated by the vertical lines). All three use $B_{\rm surf}=10^{14} \text{ G}$.}
\label{fig:lum_varytres}
\end{figure}

\begin{figure}
\centering
\includegraphics[width=0.49\textwidth,angle=0]{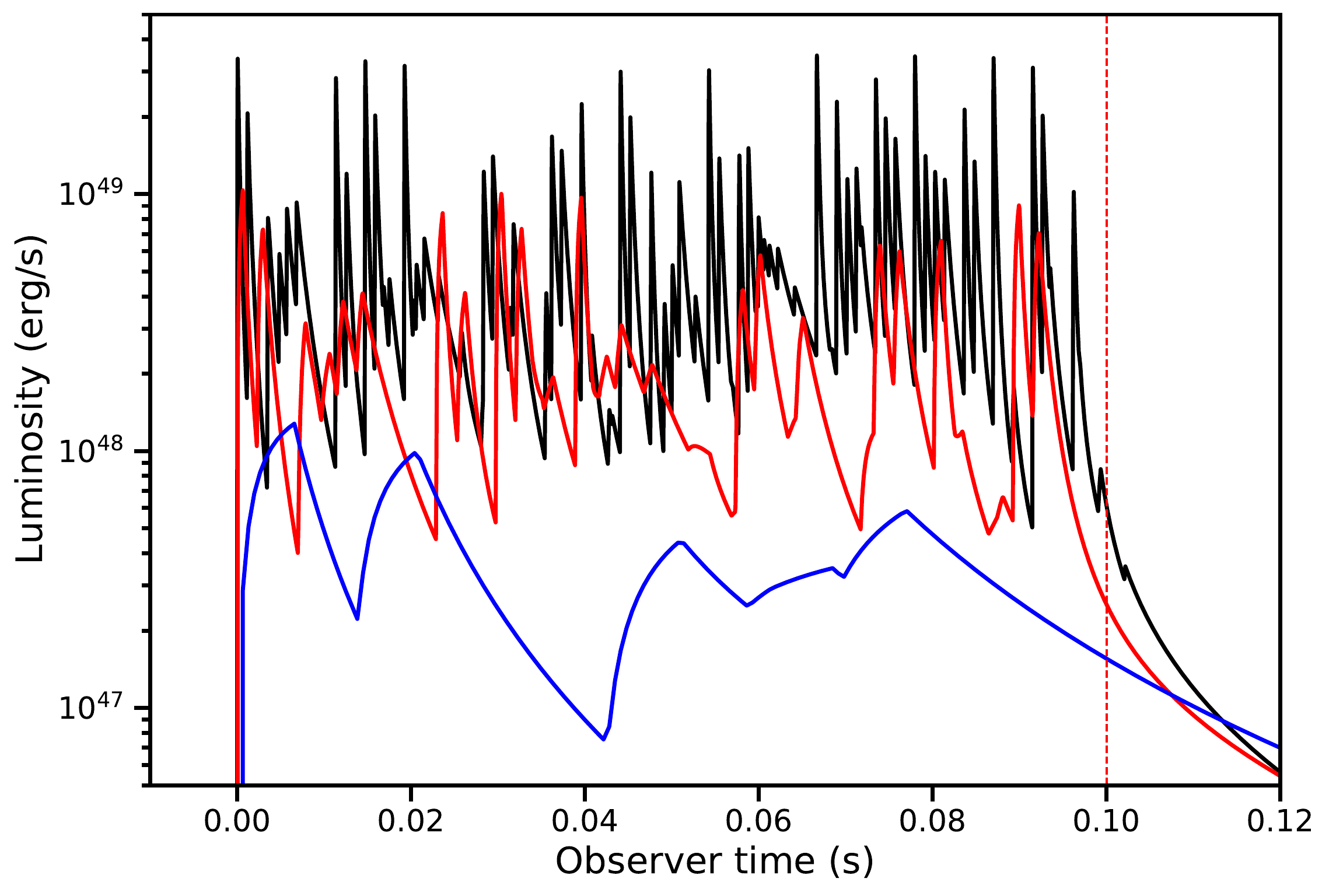}
\caption{The luminosity of the RSF produced by NSs with different magnetic field strengths. The blue, red and black lines are for $B_{\rm surf}=10^{13.5},10^{14},10^{14.5} \text{ G}$, respectively. The resonance timescale is $0.1\text{ s}$ for all three.}
\label{fig:lum_varyB}
\end{figure}

The time and radial distance at which each collision occurs are used to find the times at which their emission reaches a distant observer. The emission from each collision is then summed to obtain the total luminosity of the flare as a function of observer time. Figure~\ref{fig:lum_varytres} shows the emission observed for three different values of the resonance timescale, and we can see that the duration over which emission is observed is approximately equal to this timescale. The resonance timescale is dependent on the resonant mode's frequency and the chirp mass of the binary \citep{tsang2012resonant}, and therefore so is the duration of the flare. For the rest of this work we will assume that the flare duration $t_{\rm flare}\approx t_{\rm res}=0.1 \text{ s}$, as the detectability of a RSF will be far more dependent on its average luminosity than its duration. Figure~\ref{fig:lum_varyB} shows the impact of $B_{\rm surf}$ on RSF luminosity. Higher $B_{\rm surf}$ decreases the shell emission timescale, and thus increases the total number of shells emitted during the resonance window. A higher number of shells being emitted means that more collisions will occur, increasing the average luminosity of the flare.

\begin{figure}
\centering
\includegraphics[width=0.49\textwidth,angle=0]{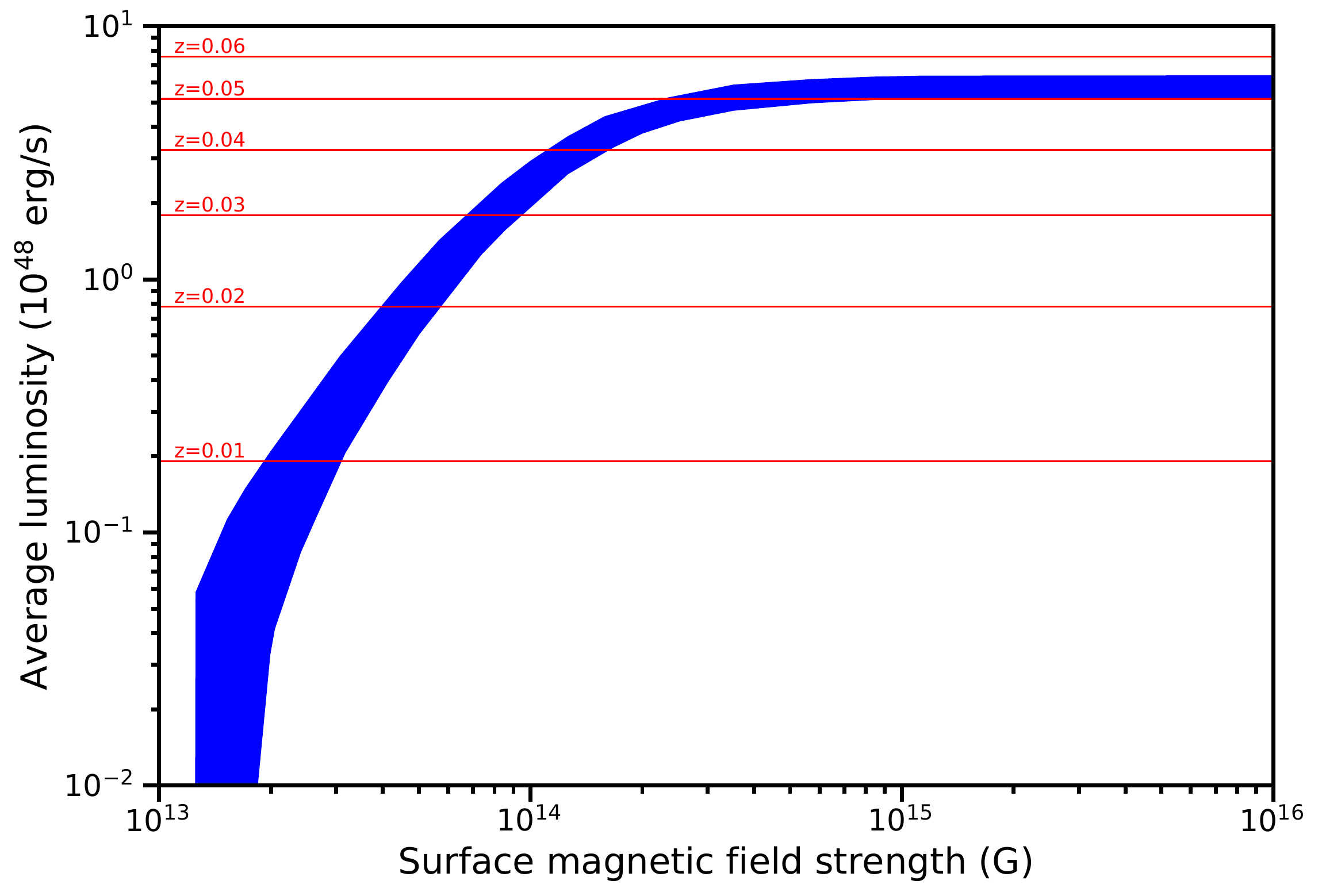}
\caption{The impact of surface magnetic field strength on the average non-thermal luminosity produced at internal shocks. The spread in luminosity is due to the randomness used in the shells' masses. The red lines are estimates of the source luminosity required for RSFs emitted by mergers at different redshifts to be detectable by \textit{Fermi/GBM} (see Section~\ref{sec:detect}).}
\label{fig:B_lum_range}
\end{figure}

In Figure~\ref{fig:B_lum_range} we show the relationship between the surface field strength and the average non-thermal luminosity of the flare during $t_{90}$ (the time period over which 90\% of the flare's energy is emitted). The spread in luminosity at any given field strength is due to the variability in the mass loading of the shells. For high field strengths (greater than $\sim10^{15} \text{ G}$) we find that the luminosity stops increasing. This is because the number of shells (equation~\ref{eq:num_shells}) becomes dominated by the breaking timescale when the emission timescale is significantly smaller, and thus further increasing the field strength (and therefore decreasing $t_{\rm emit}$) has little impact. Similarly, the cutoff at $B_{\rm surf}\approx1.2\times10^{13} \text{ G}$ is due to the emission timescale becoming long enough that only one or zero shells are emitted, and therefore no shell collisions can occur.

Whether or not a RSF is detectable is not only dependent on the total luminosity of the flare, but also on its spectrum, which determines the fraction of the flux that is in certain detector bands. The luminosity of a flare comes from emission of the internal energy that is converted from kinetic energy in order to obey conservation of momentum during shell collisions. This energy is emitted via synchrotron emission, where electrons are repeatedly excited across the shock produced in each shell during a collision. Synchrotron spectra are dependent on many parameters of the fireball shells at the internal shocks, such as their electron density, the fraction of electrons that are shock accelerated, and the fraction of their energy that is in the magnetic field \citep{rybicki1979radiative,sari1998spectra}. As these parameters are difficult to estimate, have wide reasonable ranges, and may vary between shells, we do not calculate the synchrotron spectra for each collision.

Instead, as one of the aims of this work is to asses whether RSFs are bright enough to be the observed SGRB precursor flares, we simply assume a spectrum for RSFs that peaks around the detector bands in which precursor flares are observed. 
The spectrum we choose is the Band function \citep{band1993BATSE}:
\begin{align}
N(E) = \begin{cases} AE^{\alpha}e^{-\frac{E\left(\alpha+2\right)}{E_{\rm peak}}}, & \text{if }  E\leq\frac{\left(\alpha-\beta\right)}{\left(\alpha+2\right)}E_{\rm peak} \\ A\left(\frac{\left(\alpha-\beta\right)}{\left(\alpha+2\right)}E_{\rm peak}\right)^{\left(\alpha-\beta\right)}E^{\beta}e^{\left(\beta-\alpha\right)}, & \text{if } E\geq\frac{\left(\alpha-\beta\right)}{\left(\alpha+2\right)}E_{\rm peak} \end{cases},
\label{eq:Band_func}
\end{align}
\noindent where $A$ is the amplitude, which is chosen to match the luminosity of the RSF. The parameter values that should be used in this model are not obvious, as precursor flare spectra are not well known, except that they usually peak in the same energy bands as SGRBs. However, similar to RSFs, synchrotron processes at internal shocks is a likely emission mechanism for SGRBs \citep{Sari1997Cosmological}. Within the internal shock model SGRBs and RSFs may have similar efficiencies for converting kinetic energy into radiation \citep[$\sim20$\%, e.g.][]{guetta2001efficiency}, although there is a large degree of uncertainty in the efficiency of SGRB production \citep{Granot2006Implications}. Because of these similarities we shall simply assume that RSFs have similar spectra to SGRBs and use $\alpha=-0.5$, $\beta=-2.4$ and $E_{\rm peak}=500\text{ KeV}$ \citep{nava2011spectral}, although it should be noted that variation of these parameters within reasonable ranges can easily cause the gamma-ray flux to change by a factor of a few.

\subsection{Afterglow}
The shells emitted by resonant shattering travel through the medium surrounding the merger, sweeping up matter in front of them. As the mass of the swept up matter becomes comparable to the shell mass, reverse and forward shocks form which travel through the shell and the ISM. Similarly to the shocks between shells, these shocks accelerate electrons, which emit synchrotron radiation \citep{rees1992relativistic,meszaros1997optical}. This radiation can be observed as the afterglow of the main flare, which continues to be emitted long after resonant shattering occurred.

SGRBs also have afterglows. Unlike RSFs, the highly collimated nature of SGRBs means that initially we might not see the afterglow. However, as the beam collides with the surrounding medium it spreads out, increasing the angle from which the afterglow can be seen \citep{vaneerten2011}. This results in the afterglow fluxes of off-axis events being initially low and rising over time until they peak much later than for on-axis events (see e.g. Fig. 3 of \citealt{troja2020thousand} or, more generally, e.g. \citealt{lamb2017, ryan2020gamma}). 
The mass loading of SGRB shells is expected to be significantly higher than that of RSFs, and therefore SGRB deceleration radii -- the radii by which the shells will have swept up masses comparable to their original masses -- will be correspondingly larger. As a result, we expect that RSF afterglows will peak earlier than even on-axis SGRBs. 
These effects combined mean that, while RSFs are significantly less energetic than SGRBs, at early times the afterglow of a RSF may be stronger than that of an off-axis SGRB, perhaps making it detectable. We therefore investigate the light curves of afterglows as another source of RSF emission.

After the shell collisions described in the previous Section, we are left with a small number of shells, ordered by Lorentz factor such that they will not collide without interacting with the surrounding medium. For simplicity, we assume that (before the afterglow becomes significant) all surviving shells merge into a single shell which contains the total mass and energy remaining after prompt emission. We make use of \textit{afterglowpy} \citep{ryan2020gamma} to calculate the afterglow of this single shell, assuming that it is spherically symmetric, constant in density, and that all of the matter is travelling at the same speed. \textit{afterglowpy} is a python module which uses a simple model (the dynamics of which follow \citealt{vaneerten2013, nava2013}) to calculate synthetic light curves and spectra for afterglows. This module has previously been used in \citet{troja2020thousand} to reproduce the afterglow of GRB170817A, showing that it can be a decent model for off-axis SGRBs. By comparing the afterglows calculated for SGRBs and RSFs, we can investigate whether the afterglow of a RSF could be detected, and if a SGRB from the same merger would dominate the afterglow signal, the results of which we show in Section \ref{sec:afterglow}.

\subsection{Thermal emission and optical depth}\label{sec:thermal}
So far we have assumed that coupling between seismic waves in the NS crust and the magnetic field always results in fireballs. However, a fireball can only be produced when photons are emitted from the magnetic field at a rate which makes the emission site optically-thick to pair production, otherwise photons emitted from the magnetosphere will simply escape as a dim X-ray flare, similar to magnetar flares. To check whether fireballs will be produced by resonant shattering, we follow \citet{shemi1990appearance}, who find that the emission must be launched at a temperature of at least $k_BT_0\approx 0.0164 \text{ MeV}$. For an optically-thick fireball in thermal equilibrium, the launch temperature can be approximated as \citep{goodman1986gamma}
\begin{align}
k_BT_0=0.28\left(\frac{R_{s,0}}{10^7\text{ cm}}\right)^{\frac{-3}{4}}\left(\frac{E_0}{10^{46}\text{ erg}}\right)^{\frac{1}{4}}\text{ MeV},
\label{eq:temperature}
\end{align}
\noindent where $R_{s,0}$ is the radius of a sphere with the same volume as a fireball shell immediately after it is launched, i.e. a spherical shell with width $dr\approx ct_{\rm emit}$ above $\sim10^6 \text{cm}$ (this radius is used to bridge the gap between \citet{goodman1986gamma} and \citet{shemi1990appearance}'s spherical fireball models and our fireball shells). 
$E_0\approx E_{\rm elastic}$ is the energy extracted from the shattered NS crust. If $k_BT_0> 0.0164 \text{ MeV}$ then some of this energy will be in electron-positron pairs, and the rest in photons. 
As the emission timescale, and thus the width of fireball shells, is proportional to $B_{\rm surf}^{-2}$ (equation~\eqref{eq:emission_rate}), we can use equation~\eqref{eq:temperature} to find the minimum surface magnetic field strength for pair production to be possible: $B_{\rm surf}\approx 2.6\times10^{13} \text{ G}$, which is also approximately equal to the requirement for multiple fireball shells to be launched (Figure~\ref{fig:B_lum_range}). Therefore, our estimation that non-thermal RSFs can be produced by NSs with field strengths above $\sim10^{13} \text{ G}$ holds.

In Section~\ref{sec:prompt_shell}, we assumed that all of the energy in the fireball shells is kinetic. However, if the shells become optically-thin too soon after launch, then sufficient time may not have passed for the fireball's energy to be converted into kinetic energy, resulting in a significant thermal flare. For fireball shells from a NS with $B_{\rm surf}\approx 10^{13.5} \text{ G}$, we follow \citet{shemi1990appearance} to calculate the temperature, and thus thermal emission, of the shells when they become thin to both pair production and scattering. We estimate that these shells are launched at $k_BT_0\approx0.0216 \text{ MeV}$, and become optically-thin once they have cooled to $k_BT\approx1.2\times10^{-4}\left(\frac{M}{10^{23}\text{ g}}\right)^{-\frac{1}{2}}\text{ MeV}$, where $M$ is the shell mass. By this point (if $E_0\gg Mc^2$, i.e. $M \ll 10^{25}\text{ g}$ for the shell energy we use), $\frac{E_{\rm rad}}{E_0}\approx\frac{T}{T_0}$ of the initial energy is still radiation, which escapes as a thermal flare; the remaining energy has been converted into baryon kinetic energy. For our chosen range of shell masses this corresponds to $\sim0.55-5.5\%$ of the initial energy, or $E_{\rm rad}\lesssim 10^{45} \text{ erg}$. Therefore, the resulting thermal flare has luminosity $L_{\rm BB} \lesssim 10^{47}\text{ erg/s}$, which is unlikely to be detected. NSs with stronger magnetic fields will have even less energy remaining as radiation, as their thinner shells are more dense and therefore are more efficient environments for photons to transfer energy to baryons. We therefore find that thermal emission is negligible in this fireball shell model, and so the assumption that all of the shell energy is kinetic is reasonable.

While fireball shells staying optically-thick for longer will reduce the energy lost to thermal emission, if collisions occur while they are still optically-thick, the resulting non-thermal synchrotron emission will be thermalised by scattering. To check whether this affects our model, we use the initial temperature, energy, and radius of a fireball to find the energy remaining as radiation when it becomes optically-thin, $E_{\rm rad}$ \citep{shemi1990appearance}. The radius where this occurs can then be calculated as
\begin{align}
R_{s}=R_{s,0}\frac{E_0}{E_{\rm rad}},
\label{eq:radiusandenergy}
\end{align}
\noindent where $R_{s}$ is the radius of a sphere with the same volume as a fireball shell that has just become optically-thin. For a $B_{\rm surf}=10^{13.5} \text{ G}$ NS, this is $R_{s}\approx7\times10^{10}\left(\frac{M}{10^{23}\text{ g}}\right)^{\frac{1}{2}} \text{ cm}$. Converting spheres with volume $\frac{4}{3}\pi R_{s}^3$ into shells with the same volume and width $dr\approx ct_{\rm emit}$ results in shells in our mass range becoming optically thin between $R\sim2\times10^{10}$ and $5\times10^{11} \text{ cm}$. Fireball shells from NSs with stronger magnetic fields become optically-thin at lower radii. The earliest collisions in our colliding shell model occur at distances from the NS of around $3\times 10^{12} \text{ cm}$, independent of $B_{\rm surf}$. This means that collisions occur after RSF shells become optically-thin, and therefore the non-thermal radiation will not be thermalised by scattering.

\section{Detectability of RSFs}\label{sec:detect}
While RSFs may be commonly produced in BHNS and NSNS mergers, a perhaps more important question is whether they are bright enough to be observed. \textit{Fermi}'s \textit{GBM} instrument is triggered by counts of $\sim0.7 \text{ photons/s/cm}^2$ in the $50-300 \text{ KeV}$ band \citep{meegan2009fermi}, and so we will calculate the photon flux in this range for RSF prompt emission. Unlike prompt emission, where the RSF and SGRB are well separated in time, the afterglow of a RSF will be observed at the same time as that of the main SGRB. Therefore, we compare the fluxes of these two afterglows to investigate the conditions for the RSF component to be distinguishable.

\subsection{Surface Magnetic Field Distributions}\label{sec:Bdist}
As was shown in Figure~\ref{fig:B_lum_range}, the luminosity of a RSF is strongly dependent on the NS's magnetic field strength. In order to obtain a reasonable field strength distribution, we take the period and period derivative of pulsars in the ATNF catalogue~\citep{manchester2005australia}\footnote{https://www.atnf.csiro.au/research/pulsar/psrcat/}, and calculate their surface dipole field strengths as~\citep{ostriker1969nature}
\begin{align}
B_{\rm dipole}\approx3.2\times10^{19}\text{G}\left(\frac{P}{1 \text{ s}} \cdot \Dot{P}\right)^{\frac{1}{2}}.
\label{eq:Bdipole}
\end{align}
\noindent RSFs are dependent on the total field strength at the surface, not just the dipole. The distribution of magnetic energy across the multipoles is highly uncertain, but it is known that NS  magnetic fields are far from simple dipoles \citep[see, e.g.,][]{gourgouliatos2016magnetic}. We choose to simply set to total surface field strength to a constant multiple of the dipole field strength, $B_{\rm surf} \simeq \xi B_{\rm dipole}$.

NSs with stronger magnetic fields stop being observed at younger ages than those with weaker fields, but there is no indication that these NSs stop existing. We therefore weight the contribution of each known pulsar to the overall population of NS magnetic field strengths by the inverse of the age at which it crosses the pulsar death line: the line on the $P-\Dot{P}$ plot beyond which pulsars are no longer observed. The age of a pulsar is approximated by the characteristic age \citep{ostriker1969nature}
\begin{align}
T=\frac{P}{2\Dot{P}}.
\label{eq:pulsar_age}
\end{align}
\noindent The pulsar death line varies somewhat, depending on the magnetic field structure \citep{chen1993pulsar}, NS mass and EOS \citep{zhou2017dependence}, and so on. We approximate the pulsar death line as \citep{contopoulos2006revised}
\begin{align}
\log_{10}\left(\Dot{P}\right)=3\log_{10}\left(P\right)-17.3
\label{eq:death_line}
\end{align}
\noindent and then combine equations~\eqref{eq:Bdipole},~\eqref{eq:pulsar_age} and~\eqref{eq:death_line} to find the time at which each pulsar will cross the death line if their surface magnetic fields decay in a chosen way.

The overall rate at which NS surface magnetic fields decay is uncertain. In order to cover the range of possible decays, we set an upper and lower bound. The upper bound assumes that magnetic fields do not decay (i.e. any field evolution simply transfers energy between the multipoles), that the total surface field strength is $\xi=10$ times the dipole field strength \citep[following Figure 3 of][]{gourgouliatos2016magnetic}, and that the distribution of pulsars beyond the death line is the same as the one before it.

For the lower bound, we assume that magnetic fields decay as
\begin{align}
B(t)=B_0e^{-\frac{t}{10^6\text{ yr}}}
\label{eq:B_decay_low}
\end{align}
(where $B_0$ is the NS's initial field strength), that the total surface field strength is only $\xi=3$ times the dipole field strength \citep[following the axially symmetric model in Figure 3 of][because it gives relatively weaker total fields]{gourgouliatos2016magnetic}, and that the distribution of pulsars beyond the death line is the same as the one before it. We have chosen exponential decay on a $10^6\text{ yr}$ timescale as a simple way to model fast field decay (Kostas Gourgouliatos, personal communications, 2021), but note that crustal field evolution models result in far more complex decay paths. We bin the known pulsars by their current magnetic field strengths, and weight their contributions by the ages at which they will cross the death line if their fields follow the decay given in equation~\eqref{eq:B_decay_low}. For simplicity, we ignore their current ages and assume that the observed NSs represent the birth population. With this decay, NSs reach the death line at a younger age than for the zero-decay scenario, resulting in a flatter weighting in the pulsars and therefore producing a $B_{\rm surf}$ distribution that less heavily favours high-$B_{\rm surf}$ NSs.

\begin{figure}
\centering
\includegraphics[width=0.49\textwidth,angle=0]{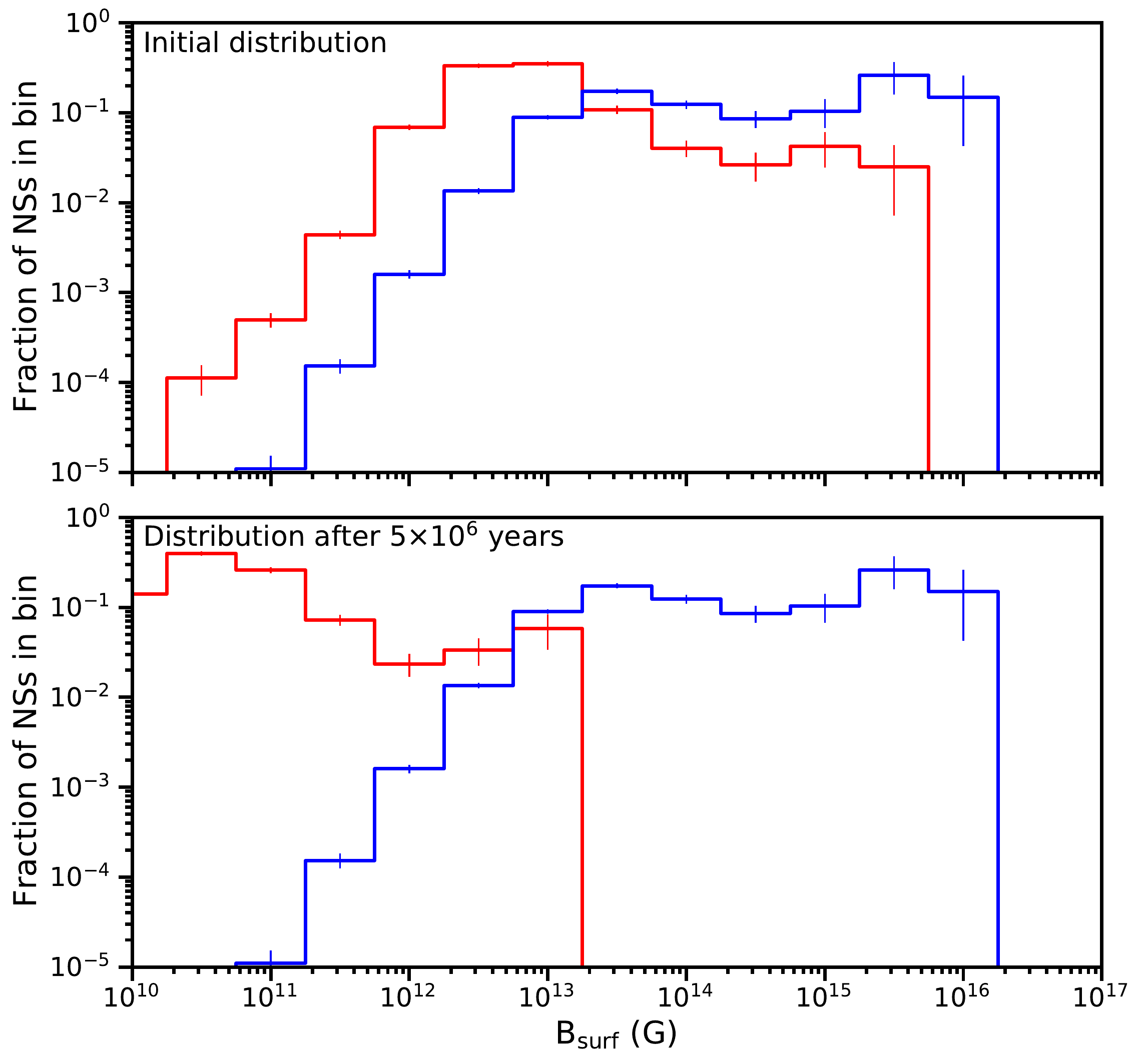}
\caption{Upper plot: the pulsar surface magnetic field distribution at the time of NS birth, showing our upper (blue) and lower (red) bounds (described in the text). Lower plot: the upper and lower bound pulsar surface magnetic field distributions after a few million years, by which point the entire lower bound has decayed below the requirement for a RSF. Both plots include $\frac{1}{N^2}$ errors, where $N$ is the number of NSs in each bin.}
\label{fig:Bbounds_binned}
\end{figure}

For both the upper and lower bounds, we ignore spun-up pulsars (which we define as $B_{\rm dipole}<10^{10}\text{ G}$), as these are a different population of NSs to those involved in compact object mergers. We also ignore NSs which are less than $10^3$ years old, as these stars may still be undergoing more complex early field and spin evolution \citep[see e.g. ][]{Gourgouliatos2019, igoshev2021evolution}, and therefore equation~\eqref{eq:Bdipole} is unlikely to accurately give their dipole field strengths. This should not significantly impact on our magnetic field distributions because the pulsar death line is located at $T>10^5$ years for all NSs in the ATNF catalogue, and therefore only a small fraction of NSs will be discounted. In Figure~\ref{fig:Bbounds_binned} we show the upper and lower bound magnetic field distributions at the time of NS formation (assuming the observed population matches the birth population), and the distributions $5\times10^{6} \text{ years}$ later. This is approximately the time by which the entire lower bound distribution has fallen below the requirement for even dim RSFs (the cutoff at low $B_{\rm surf}$ in Figure~\ref{fig:B_lum_range}, where only one shell is emitted and therefore no internal shocks form), by which point only a tiny fraction of BPASS's BHNS and NSNS systems have merged (most mergers occur after $\sim10^9$ years).

\subsection{Rate of detectable RSFs expected from BHNS mergers}\label{sec:BHNSobs}

To calculate the rate of detectable RSFs, we must first calculate the rate of BHNS (or NSNS) mergers. To do this, we again use BPASS's end products of stellar evolution and the star formation rate history and metallicity evolution described by equations~\eqref{eq:SFR} and~\eqref{eq:met_frac}. We bin cosmological redshift in steps of $dz=0.0025$ (resulting in $\sim20$ bins which are close enough for the brightest RSFs to be visible), and for each bin we calculate the number of mergers per year occurring at that redshift. Each BPASS system forms at one of 12 metallicity values, and so for each system we calculate the star formation rate at the appropriate metallicity as
\begin{align}
R_{i}(z_{\rm obs})=\left[\Psi\left(\frac{Z_{+,i}}{Z_{\odot}},z_i\right)-\Psi\left(\frac{Z_{-,i}}{Z_{\odot}},z_i\right)\right]\psi(z_i),
\label{eq:SFR_met}
\end{align}
\noindent where $z_{\rm obs}$ is the cosmological redshift value for this redshift bin, and $Z_{+,i}$ and $Z_{-,i}$ are the upper and lower ends of the metallicity bin around the $i$-th BPASS system's metallicity value. $z_i$ is the redshift corresponding to a time $\tau_0$ (see equation~\ref{eq:merge_time}) before the time corresponding to $z_{\rm obs}$, i.e. the time at which the $i$-th BPASS system would had to have formed in order to merge at $z_{\rm obs}$. 
For each of their metallicities, BPASS use a \citet{Kroupa2001IMF}-like initial mass function to calculate the fraction of star formation at that metallicity that results in each of their BHNS (or NSNS) systems (we label this fraction $\chi_i$). Using this fraction and the star formation rate $R_i$, we calculate:
\begin{align}
N_{m}(z_{\rm obs})=V(z_{\rm obs})\sum_{i}\frac{\chi_i R_{i}(z_{\rm obs})}{M_i},
\label{eq:mergers_BHNS}
\end{align}
\noindent where $N_{m}(z_{\rm obs})$ is the number of BHNS (or NSNS) mergers per year in the redshift bin at $z_{\rm obs}$, $V(z_{\rm obs})$ is the volume of that redshift bin (in $\text{Mpc}^3$), and the sum is over all of BPASS's BHNS (or NSNS) systems. Note that a division by the system's mass ($M_i$) is included in order to convert from the number of solar masses that are involved in mergers to the number of systems that merge.

We take the number of mergers in each redshift bin and use a Poisson distribution to estimate the number of mergers that occur in 10 years. For each of these mergers, we sample cut-off broken power-law fits to our upper and lower bound magnetic field distributions given in Section~\ref{sec:Bdist}. For the lower bound we use the distribution of $\tau_0$ for BPASS's systems to obtain reasonable times over which the field strengths decay before the merger. We use the resulting $B_{\rm surf}$ values with the method described in section~\ref{sec:prompt_shell} in order to obtain the source luminosities of the RSFs preceding these mergers. We convert this to the total energy flux at the observer location using
\begin{align}
L_{\rm obs}=\frac{L_{\rm source}}{4\pi D_L^2}\approx\frac{L_{\rm source}}{4\pi \left(z\frac{c}{H_0}(1+z)\right)^2},
\label{eq:lum_dist_area}
\end{align}
\noindent where we have used a low-$z$ approximation for the luminosity distance ($D_L$) as RSFs are dim and so must be at low $z$ to be detectable. To convert this to a photon flux within a particular detector band, we assume the Band function spectrum given in equation~\eqref{eq:Band_func}. 
We find the amplitude $A$ by integrating $N(E)E$ and requiring that the result be equal to the observed RSF luminosity. $N(E)$ can then be integrated over the $50-300 \text{ KeV}$ gamma-ray band to obtain the photon flux in this range. We can also invert this calculation to find the source luminosity that produces \textit{Fermi/GBM}'s required photon flux ($\sim0.7 \text{ photons/s}$) as a function of redshift, giving us an approximate source luminosity threshold for RSF observation.

\begin{figure}
\centering
\includegraphics[width=0.49\textwidth,angle=0]{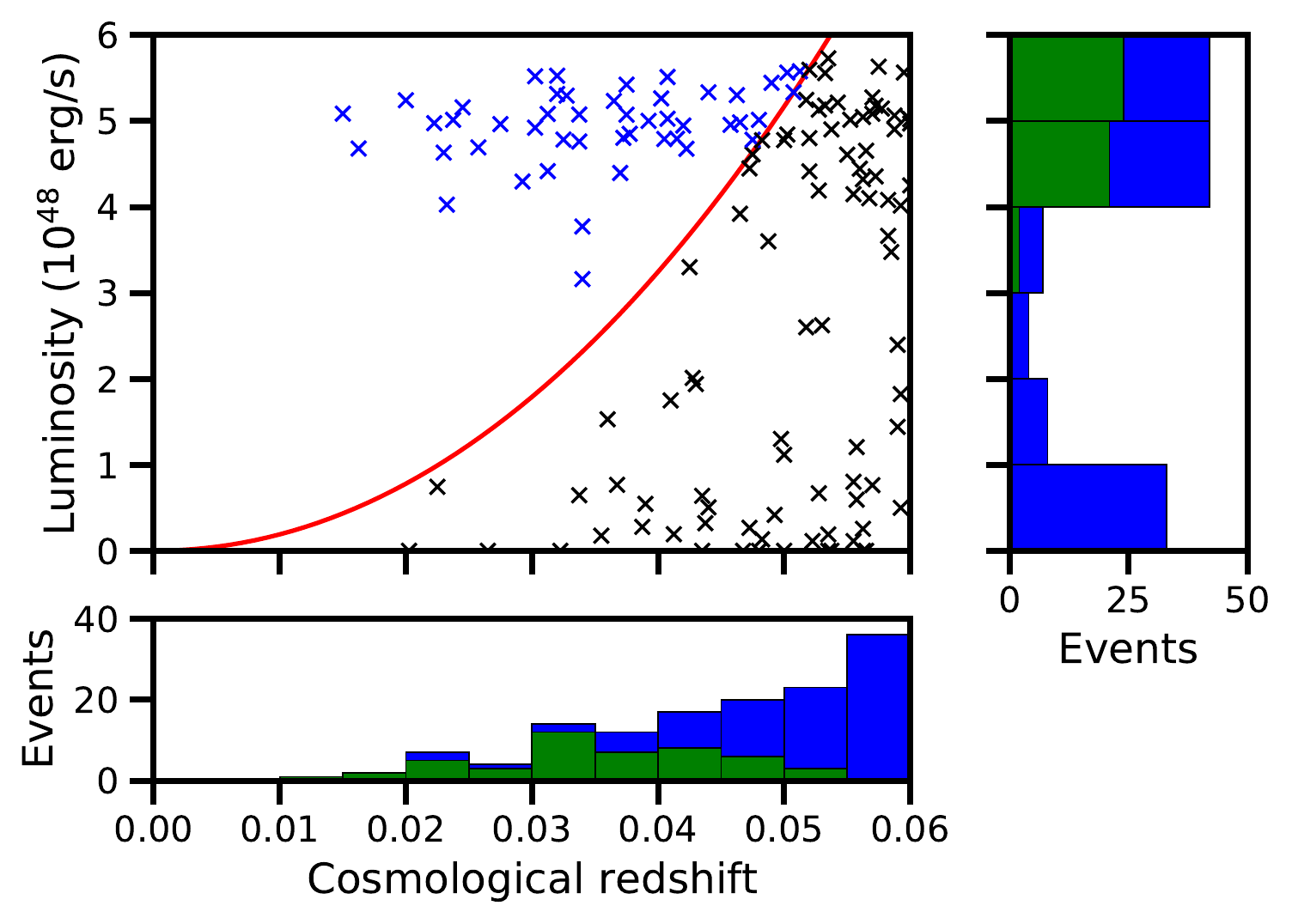}
\caption{The source luminosities of 10 years of RSFs from mergers of the BHNS systems predicted by BPASS, assuming that the NS surface magnetic field distribution at the time of the merger follows our upper bound. The red line indicates the approximate requirement for detection by \textit{Fermi/GBM}, assuming that the spectrum of all RSFs follows the Band function described in the text. The bar charts show the binned events in redshift and luminosity, with the blue bars being for all RSFs and the green for detectable ones. A luminosity of zero indicates that no RSF occurs (11 mergers). This figure shows that RSFs are visible up to $z\lesssim0.06$, and predicts that $\sim4.7$ detectable RSFs may occur per year.}
\label{fig:obs_10yr_highB}
\end{figure}

In Figure~\ref{fig:obs_10yr_highB} we show the source luminosities for a 10 year sample of RSFs produced by BHNS mergers when assuming the upper bound magnetic field distribution, alongside the approximate threshold for detection by \textit{Fermi/GBM}. Note that 11 of the events in the $0-1 \text{ erg/s}$ luminosity bin were mergers which did not produce a RSF. We see that most detectable RSFs occur at $z\lesssim0.05$, corresponding to $D_L\lesssim200 \text{ Mpc}$. While this is a very short range compared to brighter flares (such as SGRBs), it is similar to current GW detector ranges for NSNS mergers~\citep{LIGO2015advanced} and not much smaller than the range for BHNS mergers \citep{abbott2020prospects}. In this sample we find that 136 BHNS mergers occur at $z\leq0.06$ in 10 years, with 125 of them producing RSFs, 47 of which are detectable (a rate of $4.7$ per year). Note that we only consider the non-thermal emission due to collisions between shells; the thermal component is expected to be negligible.

While BHNS mergers may produce up to $4.7$ detectable RSFs per year, obtaining this many multi-messenger observations of GWs and RSFs would require 100\% detector up-time. The maximum range at which BHNS GW signals can be detected is comfortably above the maximum range for RSFs, but fluctuations in LIGO's sensitivity mean that the detection range varies over time, resulting in only a fraction of mergers within the maximum range of LIGO being observed. The simulations of \citet{abbott2020prospects} found that, for the O3 observation run, around 65\% of BHNS whose GWs were detected would occur within $\sim225\text{ Mpc}$, the maximum range for RSF detection. They also found that the time-average of the volume observed was $0.02\text{ Gpc}^3$. By multiplying this average volume by the fraction of mergers that occurred within the RSF detection range, we can find the time-averaged volume observed by GW detectors that was close enough for RSF detection to be possible, $0.02\times0.65=0.013\text{ Gpc}^3$ (assuming that the observed space has a uniform distribution of BHNS systems). As the total volume within this $\sim225\text{ Mpc}$ range is approximately $0.048\text{ Gpc}^3$, this means that only $\sim27$\% of BHNS mergers within the range of RSF detection will have detected GW signals. Our results for BHNS systems give $8.7$ BHNS mergers per year within $225\text{ Mpc}$, which corresponds to an expected value of $2.4$ BHNS mergers per year in this range with GWs detected by LIGO. Of these, we would expect $1.3$ BHNS mergers per year to have both GW detection and detectable RSFs.

The lower bound on the other hand results in an extremely low rate of detectable RSFs. The exponential decay on a million-year timescale and the lower initial $B_{\rm surf}$ distribution due to NSs crossing the death line earlier mean that even the NSs that are initially the most strongly magnetised will drop below the requirement for a RSF to occur ($\sim10^{13} \text{ G}$) after only a few million years. Therefore, only mergers that occur within a few million years of the NS's birth are capable of having any RSF, and even then these low-$B_{\rm surf}$ NSs produce dim RSFs that can't be seen from far away. The BHNS population obtained from BPASS has only $\sim0.01$\% of all mergers occurring this early, with most mergers being on a $\sim$ billion-year timescale. If we are generous and say that $\sim10$\% of all NSs have this extremely strong initial magnetisation, and that these dim RSFs are visible up to redshifts of $0.05$ (totalling $\sim8$ BHNS mergers per year), we find that detectable RSFs occur every $\sim10000$ years. Being even more generous and extending the decay timescale from $10^6$ to $10^7$ years, the fraction of BPASS's BHNS systems that merge with non-insignificant fields is $\sim1$\%, raising the rate of detectable RSFs to one every $\sim100$ years.

\subsection{Rate of detectable RSFs expected from NSNS mergers}\label{sec:NSNSobs}
We repeat the rate calculation of Section~\ref{sec:BHNSobs}, but now using the population of NSNS systems produced by BPASS. These systems contain two NSs, but we only draw one surface magnetic field strength from our distributions and therefore only allow one NS to produce a RSF. This is because accretion onto the first NS to form is likely to bury its magnetic field, reducing its surface field strength to well below magnetar levels \citep{Alpar1982, Bhattacharya1991, Cumming2001}. However, if higher multipoles are able to penetrate the accreted matter then the surface field strength may remain strong, and so two RSFs occurring is not impossible. It should also be noted that dynamically formed NSNS binaries (which we do not consider here) will not have experienced this accretion, and therefore have two opportunities to produce detectable RSFs.

\begin{figure}
\centering
\includegraphics[width=0.49\textwidth,angle=0]{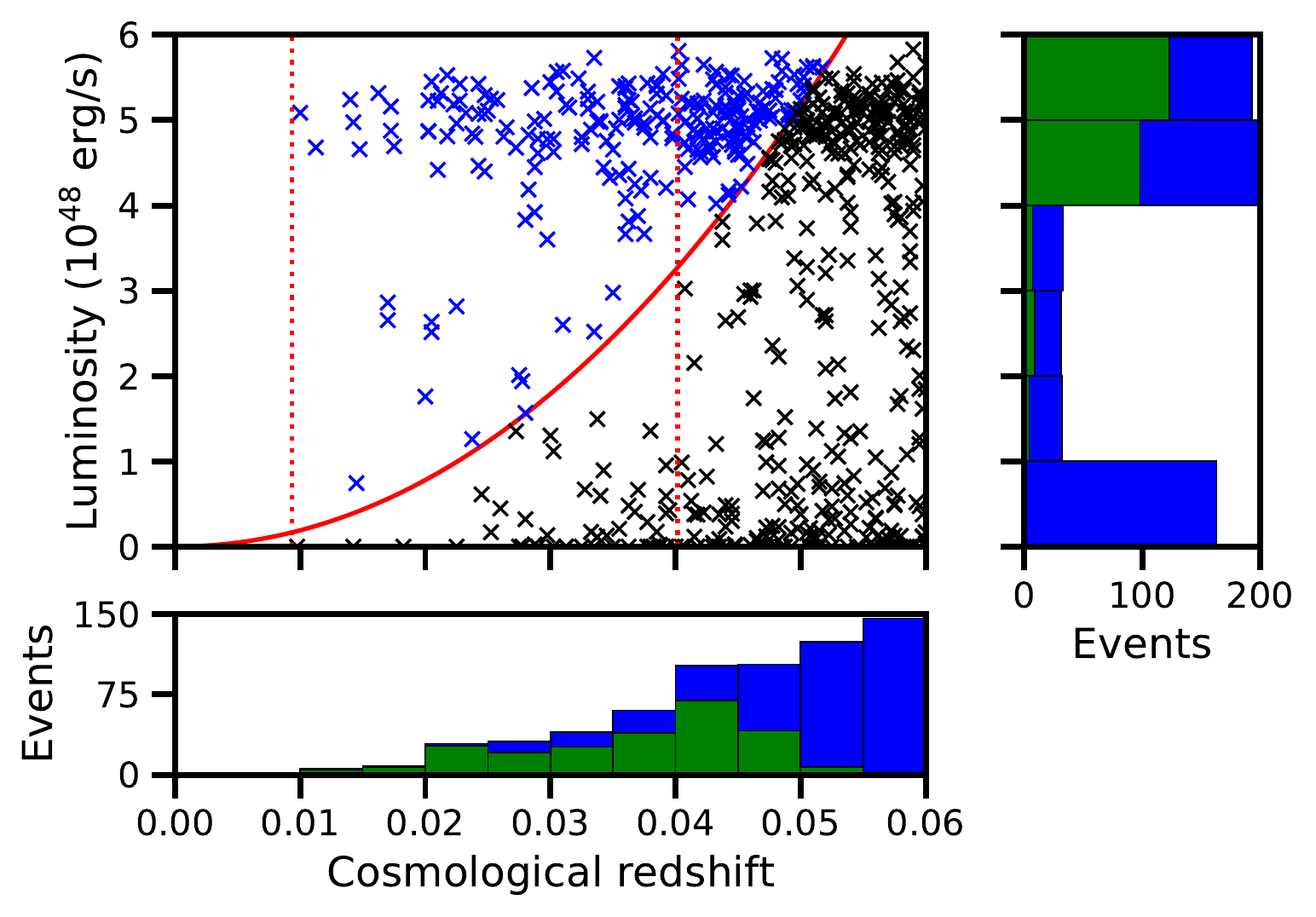}
\caption{Similar to Figure~\ref{fig:obs_10yr_highB}, but for RSFs from mergers of the NSNS systems predicted by BPASS. The red dotted lines indicate the redshifts of the two real SGRBs used in figure~\ref{fig:rsf_and_sgrb} (they are not a part of the 10-year sample and are not included in the binned events). This figure shows that RSFs are visible up to $z\lesssim0.06$, and predicts that $\sim24.3$ detectable RSFs may occur per year.}
\label{fig:obs_10yr_highB_NSNS}
\end{figure}

We obtain Figure~\ref{fig:obs_10yr_highB_NSNS} for the upper bound on the $B_{\rm surf}$ distribution. Star formation favours lower mass stars \citep{Kroupa2001IMF}, 
and therefore NSNS binaries are more commonly produced by BPASS than BHNS binaries, resulting in the rate of detectable RSFs being significantly higher. In the 10 year sample shown in the plot, $651$ NSNS mergers occur at $z\leq0.06$, with $595$ having RSFs, $243$ of which are observable ($24.3$ per year). 56 of the mergers in the $0-1 \text{ erg/s}$ bin produced no RSF. As we have assumed that there is no difference in the distribution of NS magnetic field strengths for NSNS and BHNS systems, the fractions of mergers which produce detectable RSFs are similar. The lower bound magnetic field distribution results in an extremely low rate of detectable RSFs ($\sim 0.0005$ per year), for the same reasons as for BHNS mergers.

Similar to BHNS mergers, we can not expect 100\% detector up-time and therefore the rate of multi-messenger observations will be lower than $24.3$ per year. \citet{abbott2020prospects} simulated that for the O3 observation run, around 98\% of NSNS whose GWs were detected would occur within the maximum range we found for RSF detection ($\sim225\text{ Mpc}$), and that the time-average of the volume observed was $0.0033\text{ Gpc}^3$. Assuming that the observed space has a flat distribution of NSNS systems, this means that 98\% of this volume is within the range at which RSFs are possibly detectable, $0.0033\times0.98=0.0032\text{ Gpc}^3$. Again, the total volume within this range is approximately $0.048\text{ Gpc}^3$, and therefore we would expect only $\sim6.7\%$ of NSNS mergers within $\sim225\text{ Mpc}$ to be picked up in GWs. For NSNS systems this corresponds to an expected value of $3.0$ NSNS mergers per year within $225\text{ Mpc}$ with GWs detected by LIGO, of which we would expect $1.6$ per year to also produce detectable RSFs.

\subsection{The appearance of RSFs as precursor flares}

To more clearly show what a RSF may look like in the observed photon count, we add the prompt emission of RSFs from NSs with different magnetic field strengths to the photon counts observed by \textit{Fermi/GBM} for two nearby SGRBs: GRB100216A~\citep{cummings2010short} and GRB170817A~\citep{connaughton2017fermi,goldstein2017ordinary}. We stress that there is no evidence that either of these mergers had a RSF, and that we simply use them in order to get examples of what signals which contain both RSFs and SGRBs may look like. For both of these GRBs, three of \textit{GBM}'s NaI detectors were triggered (1,2,5 for GRB170817A, and 7,9,11 for GRB100216A), and so we sum the photon counts in the triggered detectors' $50-300\text{ KeV}$ bands.

Analysis of GW170817 and GRB170817A consistently place the merger at $D_L\approx40 \text{ Mpc}$ \citep{abbott2017gw170817,abbott2017gravitational}, and so we use this value in calculating the RSF's observed luminosity. GRB100216A possibly formed within a galaxy at $D_L\approx172 \text{ Mpc}$ \citep{dichiara2020short}, and so we use this value as its distance (although this is more uncertain than GRB170817A's distance). The redshifts corresponding to these distances are shown in Figure~\ref{fig:obs_10yr_highB_NSNS}, which gives a rough idea of how bright RSFs at these distances would need to be for detection. The Band function allows us to convert the observed luminosity of a RSF ($\sim\frac{10^{48}\text{ erg/s}}{4\pi D_L^2}$) into a $50-300 \text{ KeV}$ photon flux. This flux is converted into a photon count by multiplying it by the effective area of \textit{GBM}'s NaI detectors \citep{meegan2009fermi}. We use the angles of the detectors during GRB170817A \citep{goldstein2017ordinary} to find a reasonable reduction in detector area due to their directions. This gives us an effective area of $\sim250\text{ cm}^2$, which we assume for both SGRBs. Because of the multi-messenger detection of GWs from the GW170817 merger, we can estimate the time before the merger at which a RSF would have occurred if the system had contained a magnetar~\citep{tsang2012resonant}. For the same NS EOS as used for Figure~\ref{fig:tid_dis_2.2}, a RSF would occur around $1.2 \text{s}$ before the merger, or around $3 \text{s}$ before the GBM trigger \citep[using the $\sim2.05$s time lag between the merger and trigger given by][]{abbott2017gravitational}. As there is no GW data for GRB100216A we simply use the same timings as for GW170817.

\begin{figure}
\centering
\includegraphics[width=0.49\textwidth,angle=0]{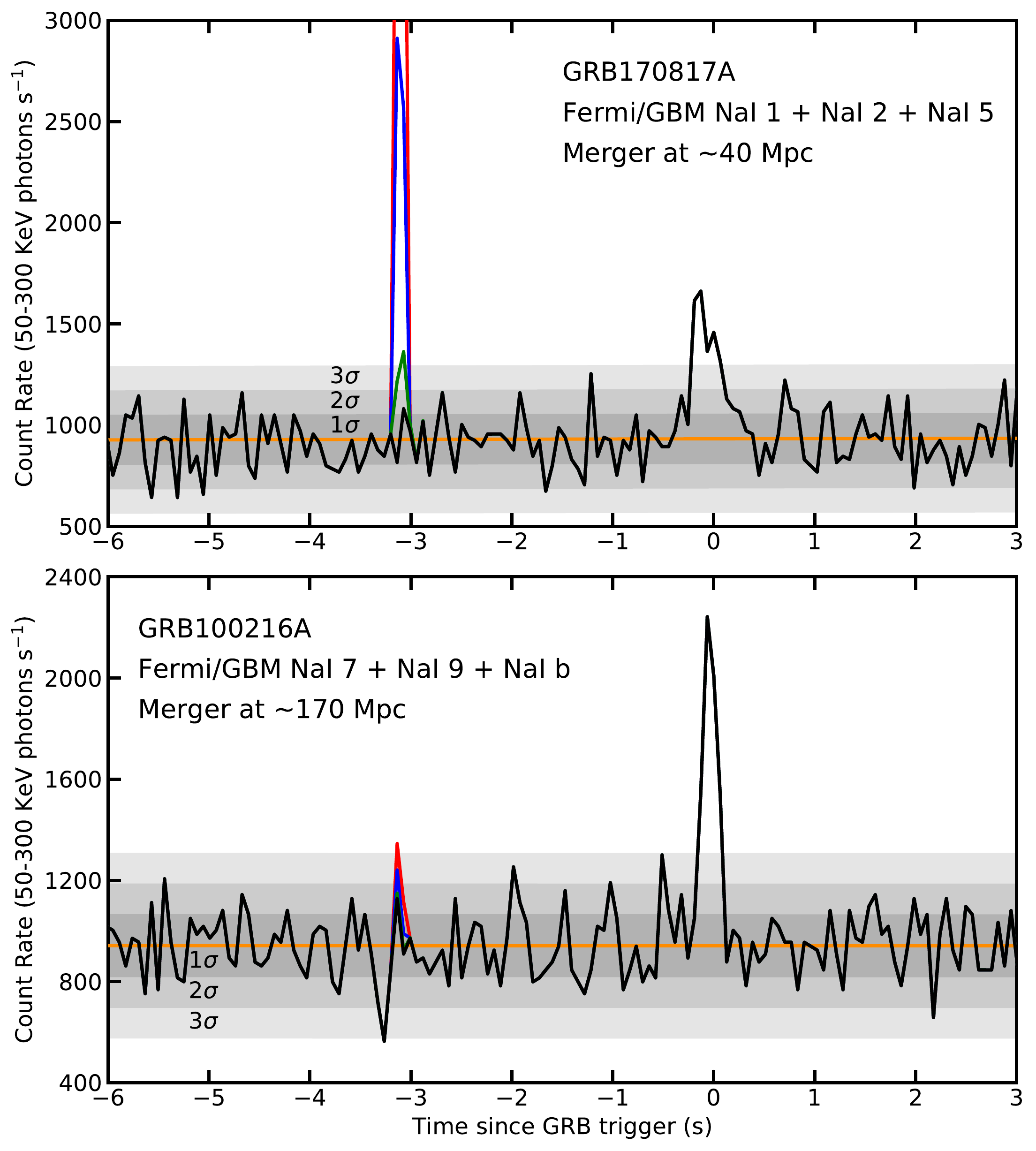}
\caption{The $50-300\text{ KeV}$ photon flux of RSFs from NSs with $10^{13.5},10^{14.0},10^{14.5} \text{ G}$ surface magnetic fields (green, blue and red, respectively), added to the photon fluxes of GRB170817A and GRB100216A detected by \textit{Fermi/GBM} (black). The $10^{14.5} \text{ G}$ NS RSF for GRB170817A (top, red) peaks at around $6000\text{ photons/s}$. Also shown is the fitted background and its first three standard deviations. The delays between these SGRBs and the mergers that triggered them are assumed to be the same (around $2.0 \text{ s}$), and the RSFs are assumed to occur around $1.2 \text{ s}$ before their mergers. We stress that neither of these SGRBs have any evidence for a RSF occurring, and that this plot is just intended to show what RSFs might look like.}
\label{fig:rsf_and_sgrb}
\end{figure}

The summed RSF and SGRB photon fluxes are shown in Figure~\ref{fig:rsf_and_sgrb}. We can see that, if one of the NSs involved in GRB170817A had been strongly magnetised ($B_{\rm surf}>10^{13.5}\text{ G}$), then a RSF would have easily been visible as a precursor flare. RSFs from NSs with weaker magnetic fields would have been difficult to discern from the background, and NSs with fields below the lower limit shown in Figure~\ref{fig:B_lum_range} are not able to produce non-thermal RSFs. Therefore, as we do not see any evidence for a RSF from GRB170817A, we can infer that the total magnetic fields at the surfaces of the NSs involved in GRB170817A were likely weaker than $\sim10^{13.5}\text{ G}$. GRB100216A however was distant enough that it would be difficult to see RSFs from NSs with all but the strongest magnetic fields, and so we are unable to put strong constraints on its NSs' magnetic fields.

\subsection{Afterglow following a RSF from a nearby NSNS merger}\label{sec:afterglow}
To get an indication of the detectability of RSF afterglows, we again investigate the flux that would have been detected if GW170817 had been preceded by a shattering event. As there is no evidence for a precursor flare being observed prior to this merger, we choose to use the strongest magnetic field strength for which a RSF would not have been visible, $B_{\rm surf}=10^{13.5} \text{ G}$ (see Figure~\ref{fig:rsf_and_sgrb}), in order to obtain an upper bound on the RSF afterglow for this merger. We will compare this to the SGRB's afterglow flux to see if a RSF's afterglow could be distinguishable, if one did occur.

\begin{figure*}
\centering
\includegraphics[width=1.0\textwidth,angle=0]{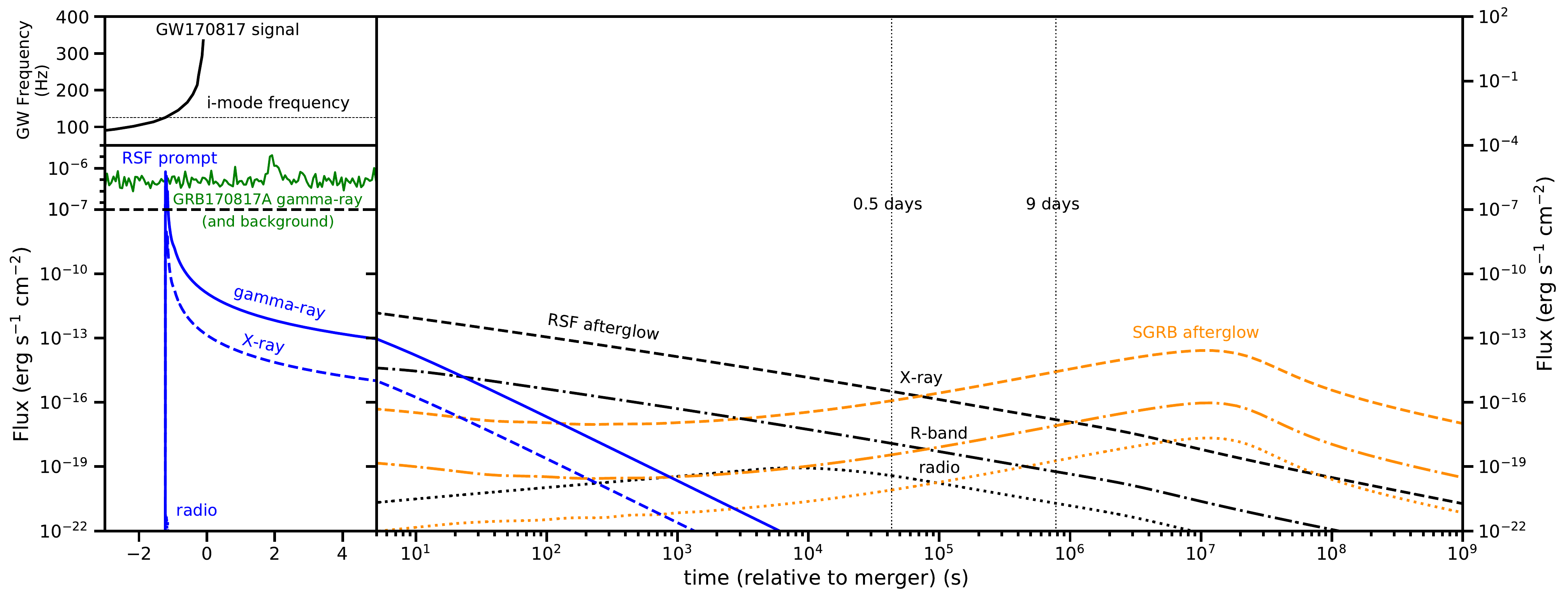}
\caption{The prompt and afterglow emission of a RSF and GRB170817A, as well as the GW170817 signal \citep{abbott2017gravitational}. In the upper left plot, we show the $i$-mode frequency for a typical NS EOS alongside the GW170817 signal, showing the approximate time at which resonant shattering occurs. The lower-left plot shows the prompt emission for the brightest RSF consistent with a lack of detection in the prompt emission (from a $B_{\rm surf}=10^{13.5} \text{ G}$ NS, with spectrum following the Band function described in Section~\ref{sec:prompt_shell}) and GRB170817A, while the lower-right plot shows their afterglows produced by \textit{afterglowpy} \citep{ryan2020gamma}. The vertical lines indicate the optimistic (0.5 days) and realistic (9 days) time required for the afterglow to be picked up. At these times, the RSF afterglow was clearly sub-dominant to that of the SGRB, and therefore would not have been identifiable. The horizontal line on the prompt indicates the transition to a linear scale to better show the GRB peak. The bands that we have plotted are: gamma-ray $=50-300 \text{ KeV}$, X-ray $=0.3-10 \text{ KeV}$, R-band $=580-720 \text{ nm}$, and radio $=1.1-3.1 \text{ GHz}$.}
\label{fig:compare_afterglow}
\end{figure*}

We use the parameters of the GRB170817A burst and jet given in \citet{troja2020thousand} as inputs for the \textit{afterglowpy} python module \citep{ryan2020gamma} to produce a fit for the afterglow of this SGRB. We use the same module to calculate the afterglow of a RSF, and assume that the dynamics of these two afterglows' emissions do not affect each other. To obtain the RSF's mass and energy, which are inputs for the afterglow model, we follow Section~\ref{sec:prompt_shell} to obtain the end state of the spherical shells after all collisions (and therefore prompt emission) have occurred. This results in $25$\% of the shells' energy being emitted in the prompt (which is shown in the lower-right of Figure~\ref{fig:compare_afterglow}), with the rest remaining as kinetic energy of the shells. A small number of shells may survive the collisions, but for simplicity we assume that all of the shells combine into a single shell with a constant Lorentz factor and density. The mass of this shell is $\sim1.6\times10^{-10} \text{ M}_{\odot}$ and its kinetic energy is $\sim1.2\times10^{47} \text{ erg}$, making the Lorentz factor $\sim400$. The X-ray, R-band and radio afterglows from the shocks that are produced when this shell collides with the ISM are shown in Figure~\ref{fig:compare_afterglow}, alongside the afterglows of GRB170817A.

In Figure~\ref{fig:compare_afterglow} we see that, by the time at which an afterglow would be found \citep[$\sim$9 days, as in][]{troja2017xray}, the afterglow of a GRB170817A-like SGRB would be similar to or stronger than that of a RSF. However, the observed afterglow of a SGRB is dependent on our viewing angle relative to its jet, with the afterglow taking longer to rise for larger viewing angles \citep{rhoads1997how,meszaros1998viewing,troja2017xray}. RSFs on the other hand are relatively isotropic, and therefore their afterglows are assumed to be independent of the viewing angle. GRB170817A was viewed from around $\sim0.4\text{ radians}$ off-axis~\citep{mooley2018superluminal,ghirlanda2019compact,troja2019year}, a relatively large angle that meant that it was only detectable due to it being close to us. For SGRBs viewed closer to on-axis (which detectable SGRBs that are further away must be), we expect their afterglows to rise at earlier times and therefore a RSF afterglow would likely be indistinguishable by the time at which the afterglow is located. However, in cases where no SGRB is produced, or the SGRB is far off-axis (and thus likely not detected), the afterglow of a nearby RSF could be visible, albeit weakly as it would peak very soon after the merger.

\section{Discussion}\label{sec:discuss}

\subsection{Neutron star properties and binary populations}

We stress that the treatment we have used for the NS magnetic field distribution is very basic. \citet{beniamini2019formation}'s more rigorous analysis of the properties of galactic pulsars found that, based on spin-down, magnetic field decay, and supernova remnants, the rate at which magnetars are born is $\sim2.3-20 \text{ Kyr}^{-1}$ (2$\sigma$ confidence), which may account for $12-100$\% of all NSs born in our galaxy. While this would appear to support our upper bound, they also found that, in order to reproduce the observed population, magnetars needed to be born with dipole fields in the range $0.3$ to $1.0\times 10^{15} \text{ G}$ which decayed over a $\sim 10^4 \text{ year}$ timescale. Such rapid decay (even faster than our lower bound) would mean that dipole fields do not survive until the merger. However, as RSFs are dependent on the total surface field and not just the dipole component, if higher order multipoles are able to survive to late times then RSFs may still be possible after the dipole has become negligible. 

Another work that investigated the birth fields of NSs is \citet{gullon2015population2}. They modelled the current populations of both radio and thermal X-ray pulsars in order to find a distribution of NSs that satisfied both. In order to reproduce the lack of observed X-ray pulsars with periods $>12 \text{ s}$, they had to restrict NS magnetic fields such that less than 1\% of all NSs were born with fields above $10^{15}\text{ G}$. This somewhat agrees with the distribution of dipole fields obtained from the ATNF catalog, but how the total field strength compares to the dipole component is still unclear. We also caution that the cutoff in X-ray pulsar periods may be due to selection effects in observation, and so the real field strength distribution could go to higher values. Together, these papers suggest that our upper bound is not wholly unrealistic as a birth distribution, but also that field decay is likely not insignificant for all NSs.

BPASS is designed for the study of the evolution of binary systems, and not the remnants left behind after the evolution. Therefore, the version of the model that we used only has a fairly simple remnant mass calculation \citep{eldridge2017binary}, and assumes a particular distribution for the kick velocity (the velocity imparted onto the remnant due to asymmetry in the supernova), ignoring the high uncertainty in this phenomena. Both the remnant masses and kick velocities are very important for calculations of compact object merger rates (kick velocities affect the binary separation and eccentricity, and thus the time for the system to merge), and therefore it is not immediately clear whether BPASS can be expected to reproduce inferred merger rates.
With the method we have used to obtain BHNS and NSNS merger rates from BPASS, we found that $\sim136$ BHNS and $\sim651$ NSNS mergers were predicted to occur at $z\leq0.06$ over a 10 year period, corresponding to $\sim191 \text{ Gpc}^{-3}\text{yr}^{-1}$ and $\sim916 \text{ Gpc}^{-3}\text{yr}^{-1}$ respectively. These rates are consistent with the inferences made by \citet{abbott2021observation} and \citet{abbott2017gw170817} after the BHNS mergers GW200105 and GW200115 and the NSNS merger GW170817. Therefore, while BPASS is not designed to study these mergers, it is not unsuitable for its use in this work. 
For a discussions on how the rates of compact mergers vary for different kick velocity and remnant mass treatments, we direct the reader to \citet{ghodla2021forward} and \citet{ Mandhai2021Zelda}.

The two EOSs that we used in Section~\ref{sec:td_tr} to find the fractions of BHNS mergers in which tidal disruption and tidal resonance occur gave significantly different results (Figures~\ref{fig:tid_dis_2.2} and~\ref{fig:tid_dis_3.0}). This is due to the $M_{\rm max}=3.0 \text{ M}_{\odot}$ EOS needing to be much stiffer than the $M_{\rm max}=2.2 \text{ M}_{\odot}$ EOS in order to reach such a high mass. A stiffer EOS gives higher radii (and therefore less compact) NSs for the same mass. As less compact NSs are more easily disrupted, the $3.0 \text{ M}_{\odot}$ EOS results in a much higher fraction of BHNS systems tidally disrupting. This same reasoning applies to the tidal disruption condition being more restrictive for higher mass NSs, as NS compactness increases significantly with mass. Tidal resonance on the other hand is more likely to occur for more compact NSs, as the orbit survives to higher frequencies. Therefore, stiffer EOSs and more massive NSs favour tidal resonance to a higher degree.

In Section~\ref{sec:BHNSobs} we found that, for our upper bound on the NS surface magnetic field distribution, around a third of BHNS mergers at $D_L\lesssim250 \text{ Mpc}$ produced detectable RSFs, while almost none were detectable when using our lower bound. As almost all BHNS experience tidal resonance, this means that anywhere from none to a third of all BHNS mergers within this distance produce detectable RSFs (depending on magnetic field evolution assumptions). 
In order to find the fraction of close BHNS mergers which are expected to produce detectable SGRBs so that we can compare it to the RSF fraction, we must combine the fraction of systems which tidally disrupt with the fraction of produced SGRBs that are detectable. 
SGRBs should be bright enough to always be detectable within $250 \text{ Mpc}$, but only if they are viewed from fairly close to the jet axis. If, as a simple example, we say that all SGRBs within this distance are visible up to $\sim0.5 \text{ radians}$ off-axis (e.g. all SGRBs have top-hat jets with isotropic equivalent luminosities $L_{\rm iso}>5\times10^{48} \text{ erg/s}$), then around $10$\% of them will be detectable. For tidal disruption, as a maximum NS mass of $3.0 \text{ M}_{\odot}$ is significantly higher than the value predicted by most astrophysical~\citep[e.g.][]{margalit2017constraining,rezzolla2018using} and nuclear physics \citep{Fattoyev:2018aa, reed2021nuclear, Ferreira2021} models, we prefer our results for the $2.2 \text{ M}_{\odot}$ EOS given in Figure~\ref{fig:tid_dis_2.2}. These results suggest that anywhere from $1.4$\% to $52.8$\% of BHNS systems experience tidal disruption, although it is likely closer to the lower end of this range as the upper bound on BH spin is extremely optimistic. Therefore, the viewing angle dependence would mean that detectable SGRBs are produced by between $\sim0.1$ and $\sim5$\% of close BHNS mergers. Comparing this to fraction of mergers which produce detectable RSFs we find that, if the NS magnetic field distribution is close to our upper bound, detectable RSFs may be $\sim10-100$ times more common than detectable SGRBs from BHNS mergers within the range of current GW detectors. This would make RSFs the dominant EM counterpart to GW detections of BHNS mergers.

\subsection{Alternate assumptions within the colliding shell model}\label{sec:shell_choices}
When calculating RSF luminosities we have assumed that, once the NS crust has shattered, tidal excitation of the $i$-mode can not occur until all of the seismic energy has been released in a fireball shell. This is because the $i$-mode primarily oscillates due to shear forces within the crust, and therefore significant disruption of the crust will likely prevent $i$-mode resonance. The crust must therefore heal before more energy can be transferred into this mode. However, if the crust were to sufficiently heal such that resonance could occur before a shell has been fully emitted, the time gap between shattering events would be reduced. This would increase the number of shells emitted during the resonance window and thus increase the average luminosity of the flare, and would slightly reduce the magnetic field strength required to produce dim RSFs.

In Figure~\ref{fig:lum_varytres} we showed that there is a clear correlation between the resonance timescale and the duration of RSFs. While in this work we have fixed it to $0.1\text{ s}$, the resonance timescale should actually be dependent on the chirp mass of the binary and the frequency of the resonant mode \citep{tsang2012resonant}. As a GW signal would provide a strong constraint on the chirp mass of a binary merger, the duration of a coincident RSF could provide an indirect measurement of the frequency of the resonant mode. This could complement the more direct measurement obtained from the GW frequency at the time of a RSF being approximately equal to the resonant (quadrupole-mode) frequency.

During a shell collision, the efficiency at which kinetic energy is converted to photons at the shocks is dependent on the ratio of the shells' Lorentz factors. It is reasonable to assume that a similar amount of energy is required to shatter the NS crust each time a shell is emitted ($\sim2\times10^{46}\text{ erg}$), and so the range of Lorentz factors is determined by the range of shell masses. The magnitudes of the shell masses are only important if they are high enough that the shells are non-relativistic or low enough that most of their energy is not kinetic: otherwise only the ratio of the masses of colliding shells affects the luminosity. For the shell energy we have used, this would require a significant fraction of the NS crust to be carried away with the shells, which seems highly unlikely (and if it were to happen, significant $i$-mode disruption may prevent further resonance). During this work, we have somewhat arbitrarily assumed that shell masses follow a flat distribution in log-space over a 2-order-of-magnitude range. The luminosity of RSFs increases by $\sim20$\% if this range is expanded by an order of magnitude, while decreasing the range by that amount would $\sim$halve the luminosity, resulting in fewer (but not zero) RSFs being detected. Alternatively, using a flat linear-space distribution over the same 2-order-of-magnitude range of masses (which may be a better choice if the source of shells' masses is rapidly replenished) would result in lower mass ratios between shells, $\sim$halving the luminosity of the flare. Of course, a flat distribution of masses is not necessarily realistic. For example, if the mass comes from shells sweeping up the medium around the binary, or if it comes from the neutron star's atmosphere and ocean and these layers are replenished on timescales longer than the resonance timescale, then we would expect the first shell to have a significantly higher mass than subsequent ones. This would result in a much brighter flare, as each of the light and fast shells collides with the more massive, slower one.

We have also assumed that the mass and energy of the shells is emitted at a roughly constant rate over the emission timescale, and therefore that the shells have roughly constant internal density. If instead the rate is not constant, for example if the mass is front-loaded or the seismic waves in the crust cause rapid variation in the surface magnetic field strength, density gradients may cause internal shocks to form within shells before they collide with other shells. This could increase the luminosity of RSFs, although these density gradient shocks would likely be weaker than the ones formed due to collisions, making their impact fairly small.

\subsection{Observed SGRB precursor flares as RSFs}
The upper bound for the NS surface magnetic field distribution (which corresponds to the scenario in which the magnetic field is frozen into the NS's core) results in a significant rate of detectable RSFs at low redshift. On the other hand, the lower bound (which is based on the rapid decay of purely crustal fields) gives an extremely low rate of RSFs. Due to the immense complexity of NS magnetic field evolution, we can not confidently make any statement as to which of these bounds is more reasonable. 
Additionally, as the pulsar emission mechanism, and therefore the cause of pulsar turn-off at the death line, is unclear, we do not know whether the weightings we have applied to the observed pulsar population are physically sound, let alone realistic.
However, as we do see precursors flares before $\sim1-10$\% of SGRBs  \citep{coppin2020identification,troja2010precursors,zhong2019precursors}, if precursor flares are indeed RSFs then this would indicate that NS magnetic fields are not just dominated by crustal magneto-thermal processes \citep{gourgouliatos2016magnetic,igoshev2021evolution}, and at least some must survive until late times. This provides some indication that our lower bound is not accurate for all NSs, and that at least some fraction of them must have their magnetic fields partially frozen into their cores \citep{Ho2017} such that they evolve over significantly longer timescales than purely crustal fields.

Assuming that some precursor flares are indeed RSFs, we can easily infer that our lower bound for the surface magnetic field distribution is much too pessimistic. However, similar inferences about our upper bound require a more detailed look at the fraction of SGRBs which are accompanied by a precursor. Our estimates for RSF luminosity require that all SGRBs with detectable RSFs must have originated at low redshift (otherwise the RSF would not be visible). For the parameters we have used, the RSF luminosity plateaus at around $5\times10^{48} \text{ erg/s}$ for NSs with surface fields of a few times $10^{14} \text{ G}$ (see Figure~\ref{fig:B_lum_range}). Flares of this strength can be seen (with e.g. \textit{Fermi}/GBM) up to redshifts of $z\lesssim0.05$ (see Figure~\ref{fig:obs_10yr_highB}) (distances of $D_L\lesssim200\text{ Mpc}$), and so all SGRBs with observed precursor flares (RSFs) must occur within this range. By comparing the fraction of observed SGRBs which have precursor flares to an estimation of the fraction of observable SGRBs that occur within this range, we are able to get some insight into the fraction of NSs that have magnetic fields strong enough to produce bright RSFs.

We can use the same BPASS population synthesis, star formation rate history and metallicity evolution described in Section~\ref{sec:BPASS} to approximate the rate of NSNS and tidally disrupted BHNS mergers (i.e. the rate of SGRBs), and how it changes with redshift. To covert this to the rate of \textit{observable} SGRBs, we integrate the logarithmic (isotropic) luminosity distribution~\citep{wanderman2015rate}
\begin{align}
\Phi(L)=\Phi_0\begin{cases} \left(\frac{L}{L_*}\right)^{-\alpha_L}, & \text{if }  L\leq L_* \\ \left(\frac{L}{L_*}\right)^{-\beta_L}, & \text{if }  L>L_* \end{cases}
\label{eq:SGRB_lum}
\end{align}
\noindent (where $\alpha_L=1$, $\beta_L=2$, $L_*=2\times10^{52} \text{ erg/s}$, and $\Phi_0$ is the normalisation constant) from $L_{\rm req}(z)$ to the maximum SGRB luminosity. $L_{\rm req}(z)$ is the minimum luminosity required for a flare to be observable at a given redshift, and is obtained with the method described in section~\ref{sec:BHNSobs} (except that for $D_L$ we properly integrate over $z$ instead of using the $z \ll 1$ approximation, as SGRBs can be visible at much higher redshifts than RSFs). With the simplifying assumption that all SGRB jets are top-hats with an opening angle distribution that is redshift independent, we calculate the cumulative fraction of observable SGRBs as a function of redshift, obtaining the fraction that occur within the distance at which RSFs are detectable.

\begin{figure}
\centering
\includegraphics[width=0.49\textwidth,angle=0]{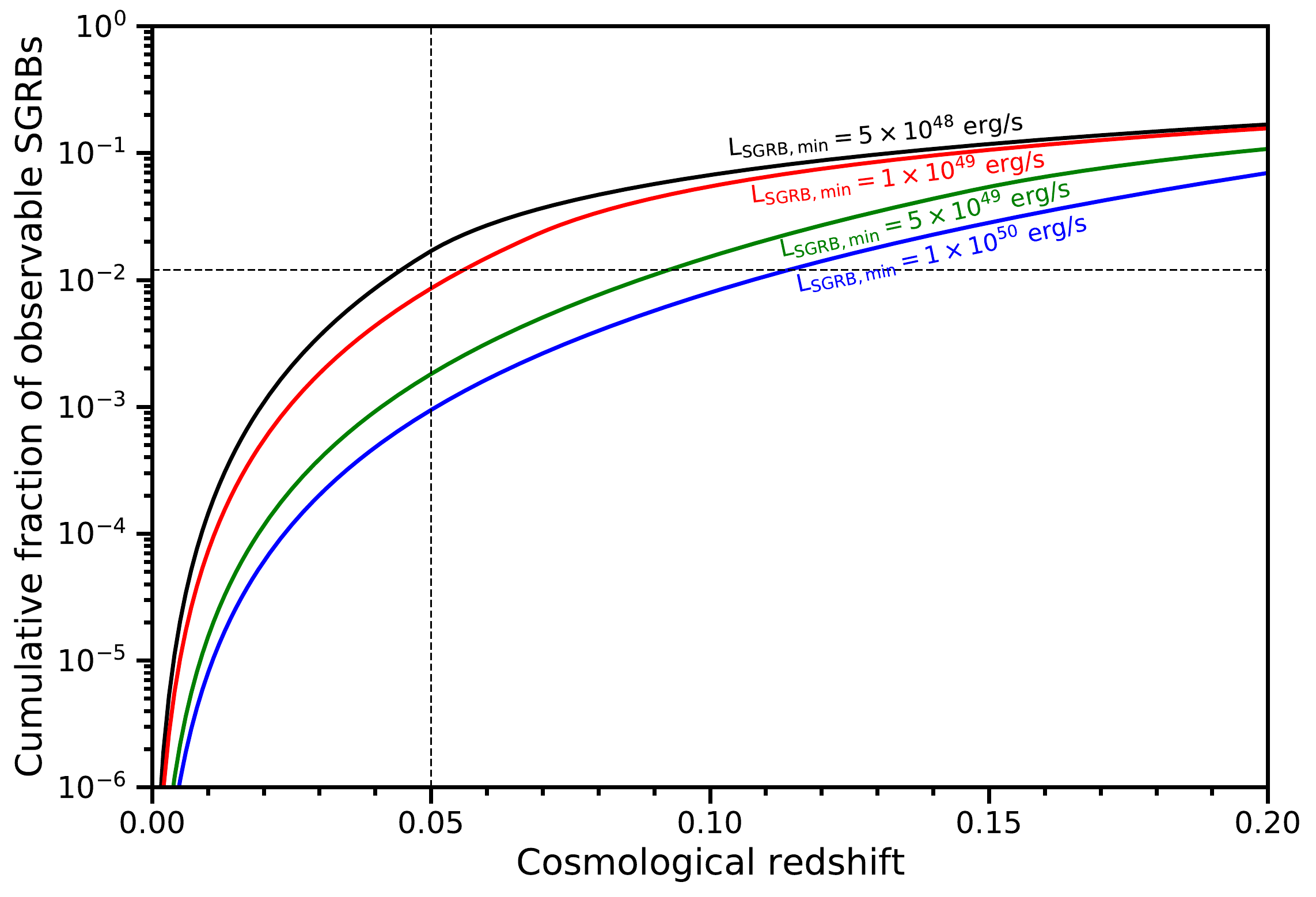}
\caption{The cumulative fraction of observable SGRBs as a function of cosmological redshift. As the low-luminosity cutoff of SGRBs is not well known, we show the result for several different values. The vertical dashed line indicates the (approximate) maximum redshift at which RSFs can be seen, and the horizontal dashed line is roughly the fraction of observed SGRBs that \citet{coppin2020identification} found had precursors.}
\label{fig:SGRB_obs}
\end{figure}

In order to normalise the SGRB luminosity distribution, we must impose maximum and minimum SGRB luminosities ($L_{\rm SGRB,max}$ and $L_{\rm SGRB,min}$). However, the behaviour of the lower end of luminosity distribution is not well known, and as $L_{\rm SGRB,min}$ has a strong impact on the fraction of observable SGRBs that are predicted to occur at low redshifts, a wide range of results can be obtained. Therefore, in Figure~\ref{fig:SGRB_obs} we show the cumulative fraction of observable SGRBs for several different values of $L_{\rm SGRB,min}$. We only use $L_{\rm SGRB,max}=10^{55}\text{ erg/s}$, as this has little impact on the results. We can see that for $L_{\rm SGRB,min}>10^{49}\text{ erg/s}$, a lower fraction of observable SGRBs occur within $z\lesssim0.05$ than \citet{coppin2020identification} found had precursor flares (see Figure~\ref{fig:SGRB_obs}). This would mean that either RSFs are brighter than we have calculated in this work, that there are other sources of bright precursors, or that some other bias exists towards observing nearby SGRBs. Lower values of $L_{\rm SGRB,min}$ on the other hand result in a higher fraction of observable SGRBs occurring at $z\lesssim0.05$ than have precursors, meaning that only a fraction of mergers need to produce bright RSFs.

For the upper bound magnetic field distribution shown in Figure~\ref{fig:Bbounds_binned}, we can see that $\sim 60$\% of NSs reach the merger with magnetic fields strong enough to produce these bright RSFs. Therefore, if $1$\% of SGRBs have observable precursors and we assume the upper bound, around $2$\% of observable SGRBs must occur close enough for RSFs to be observed ($z\lesssim0.05$). Varying $L_{\rm SGRB,min}$, we find that this corresponds to $L_{\rm SGRB,min}\approx5\times10^{48}\text{ erg/s}$ which, while lower than the values used by others \citep[see, e.g.,][and references therein]{wanderman2015rate}, is not entirely unreasonable. A magnetic field distribution below our upper bound would require a higher fraction of SGRBs to occur nearby, and therefore a lower $L_{\rm SGRB,min}$, to result in the same fraction of observed precursors. Note that it is entirely possible that the distribution of low-luminosity SGRBs requires another break in the power-law given in equation~\eqref{eq:SGRB_lum}, or that some other function should be used instead. We do not consider those possibilities here, as simply varying $L_{\rm SGRB,min}$ is enough to show that our upper bound is not excluded by the fraction of SGRBs with precursors. However, a higher fraction of SGRBs being close enough to have precursors could be explained by either a higher fraction of SGRBs being dim or off-axis, or by RSFs being brighter, and as such future constraints on the lower end of the SGRB luminosity distribution could possibly help constrain the luminosity of RSFs, if the fraction of SGRBs with precursors is also well determined. It is possible that selection biases in observation significantly alters the fraction of SGRB observed to have precursors, either favouring or disfavouring them relative to SGRBs without precursors. The $\sim 1$\% that \citet{coppin2020identification} found is lower than in previous works \citep{troja2010precursors,zhong2019precursors}, which give values up to around $10$\%, but the uncertainty in these values shows that it is currently unclear how common precursors are. Overall, the current degree of uncertainty in all the components of this calculation makes it difficult to use precursor flares to say anything concrete about RSF luminosities and the NS magnetic field distribution at this time. However, we can say that if all precursor flares are RSFs, our upper bound of little-to-no decay in NS total surface magnetic field strength is not entirely unreasonable.

Alongside RSFs, a strong NS magnetic field may trigger other EM counterparts across the spectrum, e.g. in radio and X-ray from interaction between the NS's magnetic field and its binary partner \citep{hansen2001radio,mingarelli2015fast}, in radio from shocks within magnetically driven wind \citep{sridhar2021shock}, or in orbit-modulated lensing of surface emissions \citep{schnittman2018}. Therefore, confirmation of a precursor flare being a RSF via a coincident GW detection would not only indicate that magnetic fields can become frozen into the NS core, but could also help direct the search for these other counterparts by indicating the involvement of a strongly magnetised NS in the merger.

We should also note that, while we have focused on RSFs as SGRB precursor flares, resonant shattering could produce other EM signals. For example, in the case where a weakly magnetised NS ($B_{\rm surf}\lesssim 2.6\times10^{13}\text{ G}$) experiences tidal resonance, the crust can still shatter. However, energy will not be extracted from the crust and magnetic field rapidly enough for a pair-fireball to occur. This energy may still be emitted as a dimmer magnetospheric flare (similar to crust-quake driven magnetar-flares) perhaps observable as an X-ray flare \citep[e.g. ][]{hurley1999giant,hurley2005exceptionally,jonker2013discovery}. Therefore, tidal resonance of a NS's normal modes may be a mechanism which leads to EM counterparts even in the absence of a strong magnetic field, which could be considered in the case of very nearby mergers.

\subsection{Comments on our use of GRB170817A as a fiducial SGRB}
From Figures~\ref{fig:rsf_and_sgrb} we see that, had a RSF been triggered during the GW170817 in-spiral, its prompt emission could have been easily visible, possible even outshining the main SGRB depending on the strengths of the NSs' magnetic fields. However, GRB170817A was extremely dim compared to most observed SGRBs, likely because it was viewed significantly off-axis \citep{abbott2017gravitational,troja2017xray,troja2020thousand}. If a similar event were to occur at a smaller viewing angle, the RSF (which is relatively isotropic and thus independent of the angle) would not outshine the main flare to such a degree. In contrast, a RSF at the inferred distance of GRB100216A \citep[which is similar to the lower bounds of the distances of the two recently detected BHNS mergers reported in][]{abbott2021observation} would be unlikely to outshine an observed SGRB, as even the strongest RSFs would only barely be visible.

Figure~\ref{fig:compare_afterglow} shows that the afterglow of a RSF from a $10^{13.5}\text{ G}$ NS would only be stronger than that of the SGRB for $\sim1$ day after a GRB170817A-like event, giving us a small window of time to identify the RSF afterglow, making detection unlikely. A RSF from a more strongly magnetised NS may be $\sim$ an order of magnitude brighter (afterglow flux shows a similar dependence on $B_{\rm surf}$ as the prompt luminosity does in Figure~\ref{fig:B_lum_range}), but is still likely to drop below such a SGRB's afterglow before detection. RSF afterglows may be easier to detect if the SGRB is further off-axis than GRB170817A was, as this would delay the rise in the SGRB afterglow flux. However, GRB170817A already appeared to be relatively dim due to being far off-axis, and therefore we are unlikely to detect SGRBs much further off-axis unless they are extremely close. This means that RSFs with detectable afterglows are likely to be orphan flares, for which the SGRB prompt emission is not detected (or does not occur at all, i.e. tidal disruption does not occur).

We would like to again stress that there is no evidence for a RSF occurring prior to GRB170817A. This SGRB was chosen due to it occurring at a relatively well determined distance, because said distance is low compared to other SGRBs, and because coincident GW detection is important with regards to RSFs \citep{tsang2012resonant,neill2021resonant}. The fact that no evidence of a RSF has been observed for this merger suggests that neither of the NSs involved had strong ($>10^{13.5}\text{ G}$) surface magnetic fields.

\section{Conclusion}\label{sec:conclude}

We have investigated the possibility of detecting neutron star resonant shattering flares as electromagnetic counterparts to BHNS and NSNS mergers. We compared the conditions for tidal disruption (which is required for some EM counterparts such as short gamma-ray bursts and kilonovae) and tidal resonance (which is required for RSFs), finding that tidal resonance is much more common in the population of BHNS mergers. A colliding relativistic shell model was used to calculate the luminosity of RSFs. As the number of shells emitted (and thus the luminosity of the flare) depends on the magnetic field strength at the surface of the NS, we used the properties of observed pulsars to approximate the distribution of magnetic field strengths in the total NS population. With upper and lower bounds for NS magnetic field evolution, we calculated the rate at which RSFs bright enough to be detected could be produced by BHNS and NSNS mergers. We also investigated the afterglows of these flares, finding that they could be detected if the main SGRB is very far off-axis, or if no SGRB is produced. Overall we found that if NS magnetic fields decay extremely slowly, RSF detection rates could be high. Indeed, if a future RSF is unambiguously detected prior to a merger by coincident timing of the GW inspiral signal and a gamma-ray precursor burst, then this would provide strong evidence that at least some NS magnetic fields must decay slowly.

We combined the output of the BPASS binary population synthesis code with fits for cosmic metallicity evolution and star formation history to find the population of current BHNS mergers. For this population, we showed that tidal disruption of the NS (and thus the EM counterparts that are dependent on it, such as SGRBs and KNe) is uncommon, as it can only occurs for low mass, high spin BHs with a low compactness NS partner. Tidal disruption was only found to be common if we assumed that the angular momentum transfer during accretion onto the BH was unrealistically efficient, or if we used an extremely stiff NS EOS. However, the tidal resonance condition (which determines whether resonant shattering is possible) was found to be satisfied by almost all BHNS mergers, and so it possible that RSFs will be more common counterparts to GW detections of BHNS mergers than SGRBs or KNe.

The luminosity and duration of RSFs were calculated using a model in which prompt emission is produced at internal shocks between colliding fireball shells. We found that the duration of the flare is approximately equal to the timescale over which resonant excitation occurs ($\sim0.1\text{ s}$), and that the luminosity is highly dependent on the NS's total surface magnetic field strength. For NSs with surface fields weaker than $\sim10^{13}\text{ G}$ we found that non-thermal emission is not produced as only one or zero fireball shells are emitted, meaning that only a very dim thermal RSF would be possible. On the other hand, NSs with fields stronger than $\sim10^{15}\text{ G}$ all produce similar RSFs, as the luminosity of the flare is determined by both the timescale for shells emission (which is negligible for all strong fields), and the timescale over which the crust shatters (which is independent of $B_{\rm surf}$).

To determine the rate of detectable RSFs that are produced by BHNS and NSNS mergers, we estimated the distribution of NS birth magnetic field strengths using the population of observed pulsars. We investigated two extremes for the way in which NS magnetic fields decay: an upper bound where fields experience little-to-no decay over the merger timescale (based on a field frozen into the NS's superfluid core), and a lower bound where fields undergo exponential decay over a million-year timescale (such as a field entirely contained within a thin NS crust, dominated by Hall drift and Ohmic dissipation). Using the population of binaries calculated by BPASS, we estimated the rate and distribution of ages at which BHNS and NSNS mergers occur for low redshifts. This was then combined with our magnetic field distributions and luminosity calculation to predict the luminosities of RSFs from mergers in a 10-year period. Comparing these luminosities to the required photon flux for detection by \textit{Fermi/GBM}, the upper bound field distribution gave rates of $\sim4.7$ and $\sim24.3$ detectable RSFs per year for BHNS and NSNS mergers (respectively), all of which occurred at $z\lesssim 0.05$, and around $\sim1.3$ and $\sim 1.6$ per year of which we can expect to have coincident GW detection. The lower bound on the other hand resulted in a negligible rate of detectable RSFs, as the magnetic fields decayed below the magnitude required for a RSF before all but the earliest of mergers. Therefore, if precursor flares are determined to be RSFs, a significant fraction of NSs must have their magnetic fields sufficiently frozen into the core such that the rate of decay is slow enough for their fields to last until their mergers.

Finally, we modelled the afterglow of a typical RSF in X-ray, optical (R-band), and radio bands, and compared the flux to that of GRB170817A's afterglow. The RSF afterglow flux dropped below the SGRB after only $1-2$ days, which is earlier than GRB170817A's afterglow was detected. This means that it is likely that the afterglow signal of a RSF would be lost within the afterglow of a GRB170817A-like SGRB. Additionally, GRB170817A's afterglow was relatively slow to rise compared to other SGRBs, likely due to it being significantly off-axis; the afterglow of a closer to on-axis SGRB would likely outshine a RSF even earlier. We therefore conclude that only RSF afterglows that originate from nearby mergers which do not produce SGRBs, or produce ones that are further off-axis than GRB170817A (which are unlikely to be seen), could be detected with current surveys.

In summary, we find that RSF prompt emission could be the dominant EM counterpart to BHNS mergers and a significant EM counterpart to NSNS mergers if magnetic fields at the surface of NSs decay over timescales longer than typical merger times. If RSFs are indeed found to be the main source of precursor flares, we have shown that this would have important implications for the magnetic evolution of neutron stars. While neither of the two recent GW signals thought to originate from BHNS mergers were accompanied by an EM counterpart, both were at the upper limit of the distances at which we have predicted that RSFs would be detectable, and therefore we would not expect to see one.

Relative timing of a gravitational-wave chirp and precursor flare indicating that the flare occurred before coalescence would provide unambiguous evidence that it was a RSF.
Such a coincident detection would provide a powerful probe into the nuclear physics governing the properties of the NS matter near the crust-core interface \citep{tsang2012resonant, neill2021resonant}. 
Third generation GW detectors like the Einstein Telescope or Cosmic Explorer could push detection of solar mass mergers out to redshifts of $z \sim 1-2$ \citep{ET2020, CE2017}, and may be sensitive enough to directly detect the GW phase shift due to resonant excitation of stellar modes for nearby events without need for coincident EM detection of a RSF.

%Added due to recent papers on GRB211211A's precursor
\textbf{Addendum:} During the preparation of this work, a kilonova \citep{Rastinejad2022kilonova,Troja2022submitted} was identified for the recent GRB211211A \citep{Mangan2021GRB211211A,Stamatikos2021GRB211211A}. This GRB had a precursor, and while its duration would ordinarily classify it as a long GRB, the presence of a kilonova clearly implies that it was actually a nearby SGRB with extended emission -- in addition to providing localisation and redshift. The extended emission after the main prompt \citep{Norris2006short} is consistent with the presence of a post-merger magnetar \citep{Metzger2008short}, meaning that it was likely a NSNS merger. While merger dynamics may lead to magnetic field amplification during the merger \citep[e.g.,][]{Giacomazzo2015Producing}, the presence of a magnetar does suggest that the progenitors' magnetic fields survived until the merger. Therefore, at least one of this merger's progenitors was likely a strongly magnetised NS, which would be capable of producing a RSF.

The properties of the precursor to GRB211211A \citep[duration $\sim0.19\text{ s}$, average luminosity $\sim4\times10^{49}\text{ erg/s}$, see][]{Xiao2022quasi} are not inconsistent with the properties of RSFs that we have estimated in this work. While the average luminosity is a few times higher than we have found, our calculation is a simplified order-of-magnitude approximation; reasonable changes to our assumptions such as those discussed in \ref{sec:shell_choices} could easily explain this difference. Looking at figure \ref{fig:B_lum_range}, if the luminosity plateau were to occur at higher $B_{\rm surf}$ (e.g., if the timescale for the resonant $i$-mode to shatter the crust were shorter), a RSF luminosity of $\sim4\times10^{49}\text{ erg/s}$ would suggest that this event had a NS progenitor with $B_{\rm surf}\sim5\times10^{14}\text{ G}$.

\section*{Acknowledgements}
We would like to thank Elizabeth Stanway and Jan Eldridge for useful and informative discussions about the details of BPASS, and Kostas Gourgouliatos for useful discussion about magnetic field evolution. 
DN is supported by a University Research Studentship Allowance from the University of Bath. DT additionally thanks Brian Morsony, Samaya Nissanke, and Jocelyn Read for useful discussions, as well as the 2018 Fermi Summer School, for insight into Fermi/GBM data analysis.
HJvE acknowledges partial support by the European Union Horizon 2020 programme under the AHEAD2020 project (grant agreement number 871158).
GR's research at Perimeter Institute is supported in part by the Government of Canada through the Department of Innovation, Science and Economic Development and by the Province of Ontario through the Ministry of Colleges and Universities.
WGN acknowledges support from NASA grant 80NSSC18K1019.

% The Acknowledgements section is not numbered. Here you can thank helpful
% colleagues, acknowledge funding agencies, telescopes and facilities used etc.

% %%%%%%%%%%%%%%%%%%%%%%%%%%%%%%%%%%%%%%%%%%%%%%%%%%
\section*{Data Availability}
Details of the BPASS model and the data used in this work can be found at \url{https://bpass.auckland.ac.nz/}. 
The ATNF catalogue can be found \url{at https://www.atnf.csiro.au/people/pulsar/psrcat/}. 
The tabulated EOSs are provided via \url{https://github.com/davtsang/RSFSymmetry/}. 
The \textit{Fermi/GBM} data used in Figure~\ref{fig:rsf_and_sgrb} can be found via \url{https://heasarc.gsfc.nasa.gov/cgi-bin/W3Browse/w3browse.pl}.


\bibliographystyle{mnras}
\bibliography{RSFSymmetry,newton} % if your bibtex file is called example.bib

%\begin{thebibliography}{99}
%\end{thebibliography}

%%%%%%%%%%%%%%%%%%%%%%%%%%%%%%%%%%%%%%%%%%%%%%%%%%

%%%%%%%%%%%%%%%%% APPENDICES %%%%%%%%%%%%%%%%%%%%%

%\appendix

%%%%%%%%%%%%%%%%%%%%%%%%%%%%%%%%%%%%%%%%%%%%%%%%%%

% Don't change these lines
\bsp	% typesetting comment
\label{lastpage}
\end{document}